\documentclass[fleqn,usenatbib]{mnras}

\usepackage{newtxtext,newtxmath}
\usepackage{upgreek}

\usepackage[T1]{fontenc}

\DeclareRobustCommand{\VAN}[3]{#2}
\let\VANthebibliography\thebibliography
\def\thebibliography{\DeclareRobustCommand{\VAN}[3]{##3}\VANthebibliography}

\newcommand{\Mth}{M_\mathrm{th}}
\newcommand{\qth}{q_\mathrm{th}}
\newcommand{\Mp}{M_\mathrm{p}}
\newcommand{\Hp}{H_\mathrm{p}}

\newcommand{\hp}{h_\mathrm{p}}
\newcommand{\Rp}{R_\mathrm{p}}

\newcommand{\lsh}{l_\mathrm{sh}}
\newcommand{\OmK}{\Omega_\mathrm{K}}

\newcommand{\Tosl}{T_\mathrm{OSL}}
\newcommand{\Fdep}{F_\mathrm{dep}}
\newcommand{\Fwave}{F_\mathrm{wave}}
\newcommand{\Pp}{P_\mathrm{p}}

\newcommand{\Sigmap}{\Sigma_\mathrm{p}}

\newcommand{\dphi}{\mathrm{d}\phi}

\newcommand{\cs}{c_\mathrm{s}}
\newcommand{\csiso}{c_\mathrm{s, iso}}

\usepackage{xcolor}

\usepackage{todonotes}

\usepackage{graphicx}	% Including figure files
\usepackage{amsmath}	% Advanced maths commands

\defcitealias{Cordwell2024}{CR24}

\title[Planet-induced density wave damping]{$\alpha\beta q_\mathrm{th}$-mapping of planet-induced density wave damping in protoplanetary discs}
\author[A. J. Cordwell \& R. R. Rafikov]{
Amelia J. Cordwell $^{1}$\thanks{E-mail: ajc356@cam.ac.uk},
and Roman R. Rafikov$^{1, 2}$
\\
$^{1}$Department of Applied Mathematics and Theoretical Physics, University of Cambridge, Wilberforce Road, Cambridge CB3 0WA, UK\\
$^{2}$Institute for Advanced Study, Einstein Drive, Princeton, NJ 08540, USA
}

\date{Accepted XXX. Received YYY; in original form ZZZ}

\pubyear{\the\year{}}

\begin{document}
\label{firstpage}
\pagerange{\pageref{firstpage}--\pageref{lastpage}}
\maketitle

\begin{abstract}
Planets embedded in protoplanetary discs are capable of creating a wide variety of substructures through gravitational interactions. This process is mediated through the excitation and damping of density waves which carry angular momentum across the disc. Therefore, to interpret observations of substructures, it is critical to understand the physical processes which lead to deposition of wave angular momentum to the disc fluid. In this study, we explore the relative efficiency of viscosity ($\alpha$), cooling ($\beta$), and non-linear wave evolution ($q_\mathrm{th}$) in damping planet-generated density waves. We run a large suite of hydrodynamic simulations varying viscosity, cooling timescale, and planetary mass, from which we extract radial profiles of wave angular momentum deposition. We quantify the efficiency of different wave damping mechanisms as a joint function of planetary mass, viscosity and cooling time. We find that nonlinear wave evolution leading to shock formation is typically the most important cause of angular momentum deposition, but that cooling on timescales comparable to local orbital time reaches similar levels of importance for low mass planets (sub-thermal, $q_\mathrm{th}<1$). On the contrary, linear wave damping due to viscosity is rather inefficient, requiring $\alpha \gtrsim 10^{-1.5}$ to noticeably affect damping of waves launched by thermal mass planets. Even for lowest mass planets considered ($q_\mathrm{th}=0.025$), viscosity affects wave damping only if $\alpha \gtrsim 10^{-2.9}$. Our findings could be applied to interpret observations of protoplanetary discs; they are also important for understanding wave propagation in other types of astrophysical discs.
% 250 words
\end{abstract}


%%%%%%%%%%%%%%%%%%%%%%%%%%%%%%%%%%%%%%%%%%%%%%%%%%

%%%%%%%%%%%%%%%%% BODY OF PAPER %%%%%%%%%%%%%%%%%%

%%%%%%%%%%%%%%%%%%%%%%%%%%%%%%%%%%%%%%%%%%%%%%%%%%
%%%%%%%%%%%%%%%%%%%%%%%%%%%%%%%%%%%%%%%%%%%%%%%%%%

\section{Introduction}
\label{sec:intro}

%%%%%%%%%%%%%%%%%%%%%%%%%%%%%%%%%%%%%%%%%%%%%%%%%%

A variety of different substructures --- rings, gaps, spirals, arcs, warps --- observed in protoplanetary discs by ALMA \citep{andrews_observations_2020} and direct imaging \citep{Benisty2023} may be indicative of the various ways in which young massive planets couple to the discs in which they are embedded. It has now been broadly recognized \citep{Paar2023} that such structures can naturally result from gravitational (tidal) interaction between the planets and the discs in which they form, a subject with a long history \citep{Lin1979,GT80}.

The gravitational coupling between a planet and a disc is mediated by the excitation and damping of density waves that planets drive in the disc, through a multi-stage process. First, the planetary gravity excites a density wave, endowing it with energy and angular momentum \citep{GT79}. In turn, the planet gains or loses angular momentum, which leads to its orbital evolution: orbital migration and evolution of its eccentricity and inclination \citep{GT80}. Second, the density wave propagates through the disc, carrying its energy and angular momentum, often over substantial radial distances. If it does not dissipate as it travels, the disc remains secularly in the unperturbed state \citep{Lunine1982,Greenberg1983,Papaloizou1984,Gold1989}. Only when the density wave finally starts to dissipate does its angular momentum and energy get transferred to the disc fluid. At this point, the disc surface density starts to evolve due to the injection of the wave's angular momentum, leading to the formation of gaps and other substructures \citep{Lin1986,R02b,Cordwell2024}. The wave dissipation also contributes to disc heating \citep{R16,Ziampras2020} and the excitation of vortices \citep{deVal2007,Li2009,Cimerman2023} that can serve as sites of non-axisymmetric dust accumulation.  

From this description, it is clear that understanding mechanisms through which planet-driven density waves dissipate is of critical importance in properly interpreting the substructures observed in protoplanetary discs. Over the years, several different wave damping processes have been considered. 

Non-linear effects have long been suspected \citep{Lin1986nl} to play a role in damping planet-driven density waves by turning them into shocks. The theoretical description of this process was developed in \citet{GR01} in the local approximation and in \citet{R02} in a fully global setting. That theory demonstrates that the non-linear steepening of the wave profile rapidly evolves planet-launched density waves into shocks, even when the planet mass, $\Mp$, is not very large. Dissipation at these shocks provides a robust and efficient mechanism for damping density waves. This picture has been successfully confirmed via local \citep{Dong2011b} and global \citep{Duffell2012,Cimerman2021} simulations of planet-disc interactions.

Radiative damping --- radiative losses from the parts of the disc temporarily compressed by the density wave --- has also been proposed as a linear wave-damping mechanism \citep{Cassen1996, GR01}. A linear theory of this process under the assumption of $\beta$-cooling was developed in \citet{Miranda2020I,Miranda2020II} for two different dominant cooling modes (out of the disc plane or in-plane). Simulations \citep{Zhang2020,Ziampras2023} confirmed radiative dissipation to be an efficient wave-damping mechanism under the right conditions. 

Viscosity in the disc fluid --- internal stresses typically parametrized via the $\alpha$-viscosity \citep{Shakura1973} --- is another linear process that has long been suspected to assist wave dissipation  \citep{Lunine1982}, with theoretical description of this wave damping mechanism provided in \citet{Takeuchi1996}. However, \citet{GR01} argued that viscous damping is unlikely to be competitive in the presence of non-linear wave dissipation. This conclusion has been confirmed by simulations for a limited range of parameters in \citet{Miranda2020II}.

For completeness, we also mention the possibility of magnetic stresses affecting the propagation and dissipation of planet-driven density waves \citep{Zhu2013,Lega2025}, although we will not be exploring this mechanism in the present study.

The goal of our work is to provide a comparative, quantitative assessment of the efficiency of the three key mechanisms for damping planet-driven density waves --- non-linear, radiative and viscous dissipation. We do this through a series of controlled, global numerical simulations spanning a broad range of relevant parameters, which we then analyse using physically-meaningful quantitative metrics.

Our work is structured as follows. In Section \ref{sec:methods}, we describe our disc model and, in Section \ref{sec:num_meth}, our numerical methods. In Section \ref{sec:results} we show representative results from our simulation grid, and in Section \ref{sec:parvar} quantify how wave damping varies as a function of cooling and viscosity, before using these results to understand the regions where linear damping mechanisms are or aren't important. In Section \ref{sec:discussion}, we discuss the implications of our results prior to summarising our findings, in Section \ref{sec:summary}.

%%%%%%%%%%%%%%%%%%%%%%%%%%%%%%%%%%%%%%%%%%%%%%%%%%
%%%%%%%%%%%%%%%%%%%%%%%%%%%%%%%%%%%%%%%%%%%%%%%%%%

\section{General Setup}
\label{sec:methods}

%%%%%%%%%%%%%%%%%%%%%%%%%%%%%%%%%%%%%%%%%%%%%%%%%%

We consider a planet on a circular orbit around a central object with semi-major axis $\Rp$ interacting with a 2D gaseous disc, which we represent in polar co-ordinates, $\{R, \phi\}$. The central object has mass $M_{\star}$ and the local Keplerian angular frequency for a test mass is  $\OmK \equiv \sqrt{G M_{\star}/R^3}$.

We describe gas thermodynamics using an adiabatic equation of state 
\begin{equation}
    P = \left(\gamma - 1\right) e \Sigma,
\end{equation}
where $P$ is the 2D pressure, $\gamma$ is the adiabatic index equal to $7/5$, $e$ is the internal energy of the gas, and $\Sigma$ is surface density. We additionally include a cooling term in the energy equation,
\begin{equation}
    \label{eq:cooling_prescription}
    \left(\frac{\mathrm{d} e}{\mathrm{d} t}\right)_{\mathrm{cool}} = - \frac{e - e_0}{t_{\mathrm{c}}},
\end{equation}
where $e_0$ is the initial thermal energy and the cooling timescale, $t_{\mathrm{c}}$, is parametrised by a (constant) dimensionless parameter $\beta$ defined via
\begin{equation}
    t_{\mathrm{c}} = \beta\OmK^{-1}.
\end{equation}
By varying $\beta$ we can explore a wide range of thermodynamic disc responses to an external perturbation, from locally-isothermal ($\beta\to 0$) to adiabatic ($\beta\to\infty$). 

We assume that the disc has an initially globally isothermal temperature profile, $T(R) = \mathrm{const}$, and set the $e_0$ profile accordingly. We define the scale height of the disc as $H = \csiso/\OmK$, where $\csiso$ is the initial isothermal sound speed.
We use $H$ to set the temperature indirectly by requiring that $\Hp/\Rp = \hp = 0.05$, where subscript `p' indicates a value calculated at $\Rp$. 

The background pressure in the disc is set initially by 
\begin{equation}
    P_0 = \csiso^2 \Sigma_0(R)\,,
\end{equation}
where the initial background surface density is $\Sigma_0 = \Sigma_{0,\mathrm{p}}(R/\Rp)^{-3/2}$. In the isothermal, $\beta = 0$, case this background state will eliminate the effects of the horseshoe torque on disc evolution as there will be zero gradients in temperature and vortensity. However, in the adiabatic case, where entropy must be conserved, the horseshoe torque will affect the structure of the disc \citep{paardekooper_torque_2010, tanaka_three-dimensional_2024}. It is impossible to make a choice of background gradients which would avoid activating the horseshoe torque for all values of $\beta$.

We include shear viscosity $\nu$ in our model using the standard \cite{Shakura1973} $\alpha$-parametrization:
\begin{equation}
    \nu = \alpha c_{\mathrm{s,iso}}H.
\end{equation}
Note that here and below we are using the strictly isothermal definition of $H$.

The mass of the embedded planet is expressed in terms of the thermal mass \citep{GR01},
\begin{equation}
    \Mth = \left(\frac{\Hp}{\Rp}\right)^3 M_{\star},
    \label{eq:Mth}
\end{equation}
via a dimensionless parameter $\qth$ defined as
\begin{equation}
 \qth= \Mp/\Mth. 
 \label{eq:qth}
\end{equation}
The parameter $\qth$ controls the degree of nonlinearity of disc-planet coupling: for $\qth\lesssim 1$ ('sub-thermal' $\Mp$) planet-driven density wave propagation can be considered as quasi-linear \citep{GR01,R02}, whereas for $\qth\gtrsim 1$ ('super-thermal' $\Mp$) the coupling switches to a fully nonlinear regime.

Three dimensionless parameters --- $\alpha$, $\beta$ and $\qth$ ---  characterize three different dissipative processes affecting wave propagation --- viscosity, cooling, wave damping via shocks --- and allow us to fully specify our disc-planet interaction setup.   

%%%%%%%%%%%%%%%%%%%%%%%%%%%%%%%%%%%%%%%%%%%%%%%%%%

\subsection{Density Wave Characteristics}
\label{sec:metrics}

%%%%%%%%%%%%%%%%%%%%%%%%%%%%%%%%%%%%%%%%%%%%%%%%%%

We will compare between the different simulations using the following physical variables, which are a subset of those used in \cite{cordwell_how_2025}.
We will compare the radial distribution of excitation torque, 
\begin{equation}
    \frac{\mathrm{d} T}{\mathrm{d} R} \equiv -R \oint \Sigma(R, \phi) \frac{\mathrm{d} \Phi_\mathrm{p}}{\mathrm{d} \phi} \dphi,
    \label{eq:torque}
\end{equation}
and angular momentum deposition 
\begin{equation}
    \frac{\mathrm{d}\Fdep}{\mathrm{d}R} = \frac{\mathrm{d} T}{\mathrm{d} R} - \frac{\mathrm{d} \Fwave}{\mathrm{d} R},
    \label{eq:f_dep_def}
\end{equation}
where $\Fwave$ is the amount of angular momentum carried by the density wave, calculated as 
\begin{equation}
    \Fwave \equiv \oint \Sigma u_{R} \delta u_{\phi} \dphi,
\end{equation}
where $\delta u_{\phi} \equiv u_{\phi} - (\oint \Sigma \dphi)^{-1} \oint\Sigma u_{\phi}\dphi $
as per \cite{Cordwell2024}.
Angular momentum deposition represents the amount of angular momentum moving from the wave to the secular background state of the disc. 
In order to facilitate comparisons between different planetary masses we will normalise angular momentum fluxes by 
\begin{equation}
    F_{\mathrm{J}, 0} = \left(\frac{M_p}{M_{\star}}\right)^2 \hp^{-3} \Sigma_p \Rp^4 \Omega_p^2,
\end{equation}
the classic scaling of the one-sided planet-driven Lindblad torque \citep{GT80}. As a result, $\mathrm{d}\Fdep/\mathrm{d}R$ will be expressed in units of $F_{\mathrm{J}, 0}/\Rp$.

Near the planet, angular momentum deposition may include the effects of the horseshoe (co-rotation) torque. As such we will also calculate the positions of the horseshoe region around the planet. We do so using the methods of \cite{cordwell_how_2025}, i.e. the {\sc disc\_planet\_analysis} code\footnote{\url{https://github.com/cordwella/disc_planet_analysis}}. 

Whilst we are interested in the formation of substructures we will not analyse surface density evolution directly, and will only discuss in detail the angular momentum deposition $\mathrm{d}\Fdep/\mathrm{d}R$. Angular momentum deposition can be directly connected to both the early \citep{Cordwell2024} and late stages of gap development (in Appendix \ref{sec:vss} we show that the radial profile of a gap at viscous steady state can be given in terms of angular momentum deposition function). The radial profile of $\mathrm{d}\Fdep/\mathrm{d}R$ remains approximately time-independent while a gap is shallow. Focusing on $\mathrm{d}\Fdep/\mathrm{d}R$ allows us to isolate the effects of different damping mechanisms on the density wave itself rather than, for example, the combined effect of viscosity on both the wave and the overall gap structure (via smoothing $\Sigma$ inhomogeneities). 

In Section \ref{sec:Tex} we will also briefly discuss the migration torque, i.e. the total torque that the planet exerts on the disc,
\begin{equation}
    \label{eq:summary_T}
    T = \int_0^\infty (\mathrm{d} T/\mathrm{d} R) \mathrm{d}R,
\end{equation}
and the one-sided torque, which controls gap depth
\begin{equation}
    \label{eq:summary_tosl}
    \Tosl = \frac{1}{2} \left( \left|\int_0^{\Rp} (\mathrm{d} T/\mathrm{d} R) \mathrm{d}R \right| + \left|\int_{\Rp}^\infty (\mathrm{d} T/\mathrm{d} R) \mathrm{d}R\right|\right).
\end{equation}
We will show $\Tosl$ in units of $F_{\mathrm{J}, 0}$
and the migration torque, $T$, in units of 
\begin{equation}
    T_0 = \hp F_{\mathrm{J}, 0} =\left(\frac{\Mp}{M_{\star}}\right)^2 \hp^{-2} \Sigmap \Rp^4 \Omega_\mathrm{p}^2,
\end{equation}
i.e. the expected scaling from linear theory \citep{GT80}\footnote{The 2D migration torque is not quantitively representative of the physical 3D migration torque, \citep{cordwell_how_2025}, however, as it should still follow most of the same scaling patterns, we include it in order to systematically understand the relationship between cooling and planetary mass on migration torques.}.

%%%%%%%%%%%%%%%%%%%%%%%%%%%%%%%%%%%%%%%%%%%%%%%%%%

\section{Numerical Methods}
\label{sec:num_meth}
%%%%%%%%%%%%%%%%%%%%%%%%%%%%%%%%%%%%%%%%%%%%%%%%%%

In this study, we use {\sc Athena++} \citep{stone_athena_2020} to simulate planets embedded in 2D gas discs with embedded planets obeying the viscous Navier-Stokes equations, with varied thermodynamics.

Our simulations have a spatial resolution of $N_{R} \times N_{\phi} = 1800 \times 3600$, and span from $0.2 \Rp$ to $4\Rp$ in radius (logarithmically spaced) and the full $2 \pi$ of the disc in azimuth. Our average resolution is 24 cells per scale height. Damping zones are imposed at $R < 0.28$ and $R > 3.4$. 

We represent the 2D potential of the planet using the Bessel-type potential with fixed scale height as recommended by \cite{brown_horseshoes_2024} and \cite{cordwell_how_2025}, i.e.
\begin{align}
    \label{eq:bessel_const_h}
    \Phi_{\mathrm{B}, \Hp} &\equiv - \frac{\mathrm{G}\Mp}{\Hp} \frac{\mathrm{e}^{y_\mathrm{p}}}{\sqrt{2 \uppi}} K_0 \left(y_\mathrm{p}\right),
\end{align}
where $y_\mathrm{p} = |\mathbf{R} - \mathbf{\Rp}|^2/(4 \Hp^2)$.
In practice this equation still needs to have a numerical smoothing length to be applied and so we actually use
\begin{equation}
    \label{eq:yp}
    y_\mathrm{p} = \frac{|\mathbf{R} - \mathbf{\Rp}|^2 + \epsilon^2}{4 \Hp^2}
\end{equation}
where $\epsilon = 0.1 \Hp$ is the numerical smoothing length. In Appendix \ref{sec:sec_order}, we additionally show the results obtained using the more conventional second order (or Plummer) potential
\begin{equation}
    \label{eq:plummer}
    \Phi_{2} \equiv - \, \frac{G\Mp}{\sqrt{|\mathbf{R} - \mathbf{\Rp}|^2 + r_{\mathrm{s}}^2}},
\end{equation}
where $r_{\mathrm{s}}$ is the physical smoothing length, which we set to $r_{\mathrm{s}} = 0.65 \Hp$. 

We do not include the indirect potential of the planet as it is known to have little effect on disc-planet interaction outcomes when a deep gap has not formed yet \citep{Crida2025,R2026}. Even more relevant for our present study, \citet{R2026} have shown that the inclusion of indirect potential has almost no effect on the deposition torque density, which is our key metric. 

The planetary potential is introduced over the first 10 orbits using a sinusoidal ramping function. We measure our outputs of interest at 11 orbits since the initial introduction of the planetary potential (i.e. one orbit after the ramp up has finished). Whilst this may seem like a short period of time to ensure that $\Fdep$ is constant, it was shown in \cite{cordwell_how_2025} Appendix B that this is a sufficient amount of time to develop a wave steady state with a time-converged profile of $\mathrm{d}\Fdep/\mathrm{d}R$. 

We run simulations for each possible combination of $[\beta, \alpha, \Mp]$ from the following grid
\begin{align}
    \label{eq:beta_grid}
    \beta \in \{&\infty, 1000, 300, 100, 30, 10, 3, 1, \nonumber \\
    & \, 0.3, 0.1, 0.03, 0.01, 0.003, 0.001, 0 \}, \\
    \label{eq:viscosity_grid}
    \alpha \in \{&0, 10^{-4}, 10^{-3}, 10^{-2},10^{-1}\} \quad \text{and,}\\
    \label{eq:mass_grid}
    \Mp/\Mth \in \{&0.025, 0.1, 0.25, 1, 2.5\},
\end{align}
totalling 375 simulations.

We calculate $\mathrm{d} \Fwave /\mathrm{d} R$ not from the regular disc state outputs (i.e. evaluating $\Fwave$ from the simulation data on 2D velocity and density) but directly from the divergence of the Riemann fluxes inside {\sc Athena++}, following the method of \cite{Cimerman2021}. In viscous discs the divergence of the Riemann fluxes include the viscous contribution to the flux, some of which is simply the flux that leads to standard (axisymmetric) viscous accretion. For this reason, to isolate wave-related contributions, we post process $\mathrm{d} \Fwave /\mathrm{d} R$ by removing the azimuthally-symmetric viscous contribution from this flux, i.e. 
\begin{equation}
    \frac{\mathrm{d} G_{\mathrm{sym}}}{\mathrm{d} R} = \frac{\mathrm{d}}{\mathrm{d} R} \left( 2 \pi \nu R^3 \Sigma \frac{\mathrm{d} \Omega}{\mathrm{d} R} \right),
\end{equation}
where are $\Sigma$ and $\Omega$ are taken to be their azimuthally-averaged values. This procedure allows us to retain the non-axisymmetric component of the viscous stress which should be associated with the density wave. 

Extracting the total non-axisymmetric wave flux from the fluxes used in the Riemann solver, and then subtracting the axisymmetric viscous contribution as calculated from the 2D outputs of momentum and density introduces some additional numerical error especially for $\alpha = 0.1$.
%, but is a overall a much more accurate numerical technique than evaluating $\Fwave$ from the 2D output velocity and density structure only. 
In order to reduce this numerical error, we extract $\mathrm{d}\Fwave/\mathrm{d}R$ profiles from the simulation ten times between the 11th and 12th orbit -- i.e. every tenth of an orbit. For the same reason we also apply a Savitsky-Golay filter \citep{savitzky_smoothing_1964} to the output using a width of 12 cells, approximately half a scale height, and an order three polynomial. This does not have significant impacts on our results except in the very high viscosity case.

% NB: May include in thesis but I don't think this adds to the papers story, its just a distraction
%%%%%%%%%%%%%%%%%%%%%%%%%%%%%%%%%%%%%%%%%%%%%%%%%%

\subsection{Linear Calculations}
\label{sec:lincalc}

%%%%%%%%%%%%%%%%%%%%%%%%%%%%%%%%%%%%%%%%%%%%%%%%%%

We will also briefly compare torque excitation from our simulations to linear calculations in inviscid discs using the framework of \cite{Miranda2020I} and \citet{Fair2022,Fair2025}. In brief, their method solves the linear mode equations for waves excited by planets in global discs. We adopt the same maximum number of modes  $m_{\mathrm{max}} = 80$, grid spacing and software implementation as in \cite{Miranda2020I}. As we used the Bessel-type potential for our simulations, we implemented this potential into the existing code, by replacing the softened Laplace coefficients with the equivalent new Bessel type coefficients, $b_{\mathrm{B}}^{(m)}(R/\Rp)$, defined through 
\begin{equation}
    \Phi_{m} = - \frac{G \Mp}{\Rp} b_{\mathrm{B}}^{(m)}(R/\Rp).
\end{equation}
Explicitly these coefficients can be written out as 
\begin{equation}
    \label{eq:new_laplace_coefficent}
    b_{\mathrm{B}}^{(m)}(R/\Rp) = \frac{1}{\pi} \frac{\Rp}{\Hp} \int_0^{2\uppi} \cos(m \phi) \frac{\mathrm{e}^{y_\mathrm{p}}}{\sqrt{2 \uppi}} K_0 \left( y_\mathrm{p} \right) \mathrm{d}{\phi}\,, 
\end{equation}
where $y_{\mathrm{p}}$ is defined as in equation (\ref{eq:yp}).

We do not, at present, have a combined cooling and viscous model for the linear evolution of planet-driven density waves, and therefore chose to only compare our simulations with linear calculations in the inviscid case. We will only compare linear calculations to the torque excitation profiles from simulations, but as we will show, torque excitation is not strongly dependent on viscosity and experiences only very small non-linear effects for $\Mp < \Mth$.

%%%%%%%%%%%%%%%%%%%%%%%%%%%%%%%%%%%%%%%%%%%%%%%%%%
%%%%%%%%%%%%%%%%%%%%%%%%%%%%%%%%%%%%%%%%%%%%%%%%%%

\section{Representative results}
\label{sec:results}

%%%%%%%%%%%%%%%%%%%%%%%%%%%%%%%%%%%%%%%%%%%%%%%%%%

%%%%%%%%%%%%%%%%%%%%%%%%%%%%%%%%%%%%%%%%%%%
\begin{figure}
    \centering
    \includegraphics[width=\linewidth]{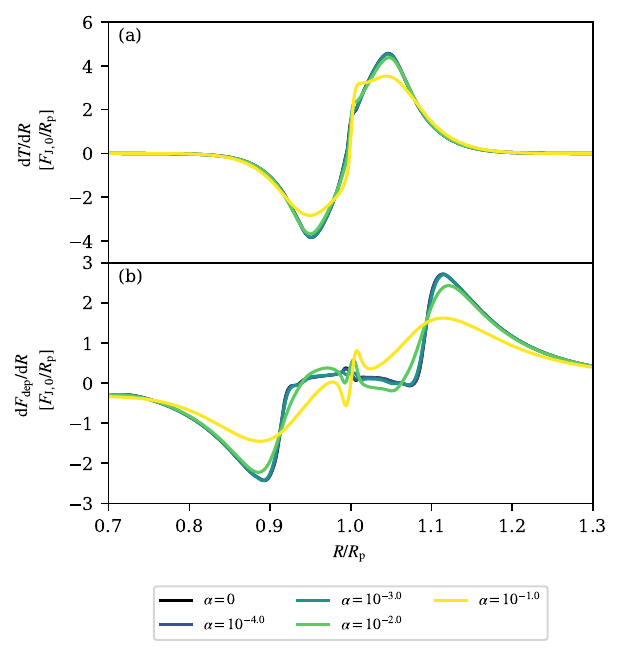}
    \caption{A representative example of (a) excitation torque density, $\mathrm{d}T/\mathrm{d}R$, and (b) angular momentum deposition, $\mathrm{d} \Fdep/\mathrm{d}R$, from isothermal simulations with a $1 \Mth$ planet ($\qth=1$), for different values of viscous $\alpha$ (indicated in the legend). The observed torque excitation is due to Lindblad resonances and angular momentum deposition is driven essentially by shock dissipation only (the wave shocks first at radii where $\mathrm{d} \Fdep/\mathrm{d}R$ starts rapidly growing in amplitude, at $R\approx 0.92\Rp$ and $1.08\Rp$). Viscosity promotes earlier angular momentum deposition, although it does not have significant effects on torque excitation or angular momentum deposition for $\alpha \lesssim 0.1$. }
    \label{fig:example_phenomology_alpha}
\end{figure}
%%%%%%%%%%%%%%%%%%%%%%%%%%%%%%%%%%%%%%%%%%%

%%%%%%%%%%%%%%%%%%%%%%%%%%%%%%%%%%%%%%%%%%%
\begin{figure}
    \centering
    \includegraphics[width=\linewidth]{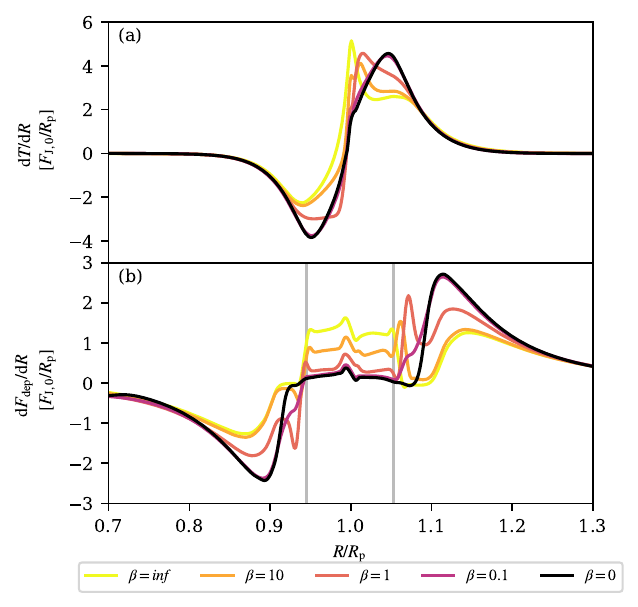}
    \caption{Representative radial profiles of (a) torque excitation and  (b) angular momentum deposition from inviscid simulations with a $1 \Mth$ planet ($\qth=1$), for different values of the cooling timescale $\beta$ (indicated in the legend). Panel (b) also shows (vertical gray lines) the radial extent of the co-orbital region occupied by horseshoe orbits as measured from the $\beta = 0$ simulations. Changes in cooling timescale modify the overall magnitude of torque excitation as well as causing to the excitation of the co-rotation torque leading to deposition in the co-orbital/horseshoe region. While in the adiabatic and globally isothermal extremes angular momentum deposition outside the horseshoe region is driven by shock formation, at intermediate values of $\beta$ there are additional extrema of $\mathrm{d}\Fdep/\mathrm{d}R$ just outside the horseshoe region caused by radiative damping of the density wave.}
    \label{fig:example_phenomology_beta}
\end{figure}
%%%%%%%%%%%%%%%%%%%%%%%%%%%%%%%%%%%%%%%%%%%

In this section, we describe representative results from a handful of our simulations. Our goal here is to identify and describe the characteristic patterns of how viscosity and cooling impact torque excitation and angular momentum deposition. In order to clearly describe how different features in torque excitation and angular momentum deposition depend on viscosity and cooling, we first consider only a single planetary mass $\Mp=1\,\Mth$ (i.e. $\qth=1$) and vary $\alpha$ (Section \ref{sec:alp}) and $\beta$ (Section \ref{sec:beta}), seperately.

%%%%%%%%%%%%%%%%%%%%%%%%%%%%%%%%%%%%%%%%%%%%%%%%%%

\subsection{Effect of varying $\alpha$ (viscosity)}
\label{sec:alp}

%%%%%%%%%%%%%%%%%%%%%%%%%%%%%%%%%%%%%%%%%%%%%%%%%%

In Figure \ref{fig:example_phenomology_alpha}, we show how the excitation torque density $\mathrm{d}T/\mathrm{d}R$ (panel (a)) and angular momentum deposition $\mathrm{d}\Fdep/\mathrm{d}R$ (panel (b)) caused by a $\qth = \Mp/\Mth=1$ planet vary with viscosity, $\alpha$, in an isothermal, $\beta= 0$, disc. We find, that with the exception of $\alpha = 0.1$, that viscosity has only negligible effects on torque excitation.

We now describe angular momentum deposition in the isothermal inviscid case, shown in the the black curve in Figure \ref{fig:example_phenomology_alpha}(b), in order to provide a point of comparison to those simulations including viscosity and cooling. In this case there is essentially no angular momentum deposition until a radial distance of a `shocking length' \citep{GR01, R02}
\begin{equation}
\lsh \approx 0.8 \Hp \left(\frac{\gamma + 1}{12/5} \frac{\Mp}{\Mth}\right)^{-2/5}
\label{eq:lsh}
\end{equation}
away from the planet is reached. Beyond that point the wave shocks and the amount of wave angular momentum deposited into the disc per unit radius steeply rises, reaches a maximum and then slowly decays. In the inner disc, not shown on the plot, there is an additional small peak in $\mathrm{d}\Fdep/\mathrm{d}R$ due to the formation of a secondary spiral arm \citep{Fung2015,Bae2018,Miranda2019I} which can cause additional gap formation \citep[as in][]{dong_multiple_2018}. 

Figure \ref{fig:example_phenomology_alpha}(b) clearly shows that the general effect of viscosity is to smooth out the features of the angular momentum deposition, lowering the maximum value of $\mathrm{d}\Fdep/\mathrm{d}R$. High viscosity also makes $\mathrm{d}\Fdep/\mathrm{d}R$ non-zero close to the planet where the shock has not yet formed, which will affect the shape of a gap that will form around the planetary orbit, see Section \ref{sec:obs}. At the same time, for the adopted $\Mp$ and $\beta$ a significant modification of $\mathrm{d}\Fdep/\mathrm{d}R$ appears only for $\alpha=10^{-2}$, while for $\alpha\leq 10^{-3}$ the deposition curves are essentially indistinguishable from the inviscid curve. Viscosity has an even weaker effect on torque excitation, with $\mathrm{d}T/\mathrm{d}R$ different from the inviscid case only for $\alpha=0.1$ (with a reduced amplitude and somewhat shifted towards $\Rp$), see Figure \ref{fig:example_phenomology_alpha}(a). 

% \We find, that for these values of $\Mp$ and $\beta$ and  for a viscosity $\alpha\lesssim 10^{-2}$, that viscosity does not have a strong effect on either torque excitation or angular momentum deposition, acting only to slightly smooth out the deposition structure for $\alpha = 10^{-2}$. 

%%%%%%%%%%%%%%%%%%%%%%%%%%%%%%%%%%%%%%%%%%%%%%%%%%

\subsection{Effect of varying $\beta$ (cooling)}
\label{sec:beta}

%%%%%%%%%%%%%%%%%%%%%%%%%%%%%%%%%%%%%%%%%%%%%%%%%%

In Figure \ref{fig:example_phenomology_beta} we show how $\mathrm{d}T/\mathrm{d}R$ and $\mathrm{d}\Fdep/\mathrm{d}R$ are affected by the variation of the cooling time $\beta$, keeping $\Mp$ and $\alpha=0$ fixed. 

As one can infer from panel (a), cooling has a significant effect on the excitation torque. The first noticeable difference between $\mathrm{d}T/\mathrm{d}R$ profiles at different values of $\beta$ is an overall re-scaling of the torque structure as expected from linear theory \citep{Miranda2020I}, which is also discussed in Section \ref{sec:Tex}. We also observe changes in the radial structure of $\mathrm{d}T/\mathrm{d}R$, with a positive co-orbital (horseshoe) torque contribution appearing right near $\Rp$ for non-zero $\beta$ and growing in strength as $\beta\to\infty$. This torque is expected to be excited only when there is a gradient in both vortensity and either temperature (for isothermal discs) or entropy (for adiabatic discs) \citep{paardekooper_torque_2010, tanaka_three-dimensional_2024}, and to be centred closely around $\Rp$. Due to our choice of background gradients (with $T(R)=$const) this torque is inactive in the isothermal case, $\beta = 0$, and the co-orbital contribution to $\mathrm{d}T/\mathrm{d}R$ is absent. In the adiabatic case, $\beta = \inf$, the horseshoe torque is active, owing to the non-zero background gradient in entropy. At intermediate values of $\beta$ there seems to be a smooth variation in the strength of horseshoe torque excitation, and while the width of this feature often exceeds that of the adiabatic case, the maximum magnitude never does, in agreement with \citet[][see their Fig. 5c]{Miranda2020I}. 

At the same time, the peak/trough of $\mathrm{d}T/\mathrm{d}R$ at $R\approx \Rp\pm\Hp$, which are associated with torque excitation at Lindblad resonances \citep{GT79,GT80}, steadily go down in magnitude as $\beta$ varies from $0$ to $\infty$. This shows that Lindblad torque excitation, which directly contributes to the angular momentum of the planet-driven density wave, is a decreasing function of $\beta$. This again is in agreement with \citet[][see their Fig. 5b]{Miranda2020I}. 

Moving on to panel (b) of Figure \ref{fig:example_phenomology_beta}, one can see that cooling also causes very significant changes in the angular momentum deposition profile $\mathrm{d}\Fdep/\mathrm{d}R$. We can break down these changes into the differences due to the excitation of the horseshoe torque and changes in the damping of the density wake (excited by the Lindblad torque). The deposition due to the horseshoe torque is roughly constant within the radial range occupied by the horseshoe orbits, which indicated with the grey lines as measured from the local streamlines in our simulations. The magnitude of this deposition structure starts from $0$ in the isothermal, $\beta = 0$, case and increases with $\beta$ up to a maximum in the adiabatic, $\beta = \inf$, case, likely owing to the increasing importance of entropy to the nature of the flow as the cooling timescale is increased. Over long time periods this co-orbital feature should disappear as the horseshoe torque saturates. 

Outside of the horseshoe region we also see important changes in the deposition of angular momentum related to dissipation of the planet-driven density wave. In both the globally isothermal and adiabatic cases, angular momentum deposition due to damping of the density wave is driven solely by the dissipation at the shock\footnote{In the absence of nonlinear shock dissipation in the adiabatic or globally isothermal cases the wave would preserve its angular momentum flux $\Fwave$, and the wave-related contribution to $\mathrm{d}\Fdep/\mathrm{d}R$ would be absent. In a locally isothermal disc where $\cs$ is a function of $R$, $\Fwave/\cs^2$ would be conserved instead  \citep{Miranda2020I}.} into which the wave evolves nonlinearly. In the adiabatic case the total amount of angular momentum deposited is smaller owing to the weaker overall Lindblad torque excitation, and deposition starts further away from the planet\footnote{This may seem unintuitive given that the standard formula (\ref{eq:lsh}) for shocking length suggests $\lsh$ increasing as $\gamma \rightarrow 1$, i.e. in the isothermal approximation. However, in the standard form of equation (\ref{eq:lsh}) the scale heigh is defined through $c_{\mathrm{s, adi}} = \gamma^{1/2}c_{\mathrm{s, iso}}$ and is equal to $\sqrt{\gamma}H$ in the notation of this paper. Accounting for this difference we find that the predicted $\lsh$ in an adiabatic disc is $1.3$ times that of an equivalent isothermal disc, which matches the changes we see in our simulations.}. But in both cases the deposition pattern in simple --- a single peak/trough in the outer/inner disc slightly beyond $\lsh$ from the planet --- and follows the description of the inviscid case in Section \ref{sec:alp}.

At intermediate values of $\beta$ situation becomes more interesting. 
For $\beta\sim 1$, one can see an additional sharp peak/trough of $\mathrm{d}\Fdep/\mathrm{d}R$ forming in the outer/inner disc just outside the horseshoe region. These features emerge before the wave has had a chance to shock: they are located at $R = 0.93 \Rp$ ($R = 1.07 \Rp$) in the inner (outer) disc, prior to the wake forming a shock which causes the main trough (peak) in deposition at around $R= 0.88\Rp$ ($R = 1.15 \Rp$). Thus, they must be unrelated to the nonlinear wave evolution and are caused by $\beta$-cooling, which is a linear process, very efficient for $\beta\sim 1$ \citep{Miranda2020I,Miranda2020II}. These features are also visible in some form for $\beta=0.1$ and 10 in Figure \ref{fig:example_phenomology_beta}(b). As some dissipation of the wave flux is accounted for by these features, the amplitude of the main deposition torque features (caused by shock dissipation) decreases. To summarize, for $0.1\lesssim \beta\lesssim 10$ linear (cooling) and nonlinear (shock) processes provide comparable contributions to density wave damping, a conclusion further corroborated in Section \ref{sec:parvar}.

%%%%%%%%%%%%%%%%%%%%%%%%%%%%%%%%%%%%%%%%%%%
\begin{figure}
    \centering
    \includegraphics[width=\linewidth]{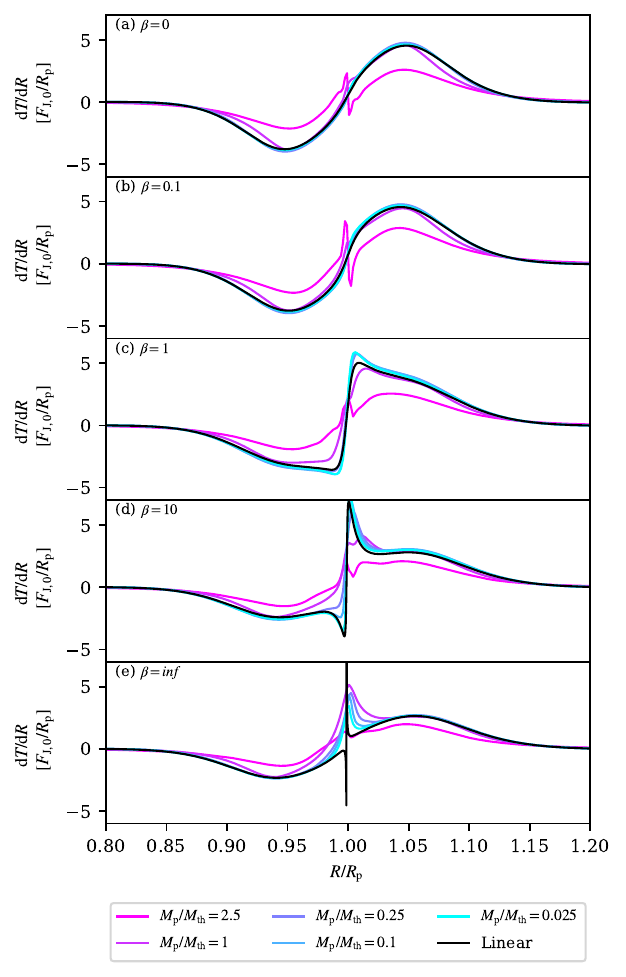}
    \caption{Excitation torque density $\mathrm{d}T/\mathrm{d}R$ as a function of planetary mass $\Mp$ (colors, shown in legend) and cooling timescale $\beta$ from inviscid simulations and computed using linear theory (black). Note differences from linearity for both the co-rotation torque and overall torque excitation. Linear theory accurately captures torque excitation due to Lindblad torques in the low mass cases ($\Mp\lesssim 1\,\Mth$) and for $\beta\leq 10$. See text for details.}
    \label{fig:torque_comparison}
\end{figure}
%%%%%%%%%%%%%%%%%%%%%%%%%%%%%%%%%%%%%%%%%%%

%%%%%%%%%%%%%%%%%%%%%%%%%%%%%%%%%%%%%%%%%%%
\begin{figure}
    \centering
    \includegraphics[width=\linewidth]{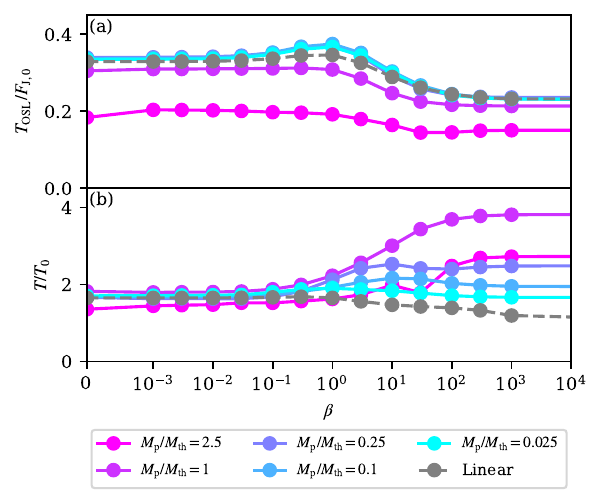}
    \caption{
    (a)  One-sided ($T_\mathrm{OSL}$) and (b) total ($T$) torques compared with linear theory as a function of cooling time $\beta$ and planetary mass $\Mp$ (colors, see legend). Linear calculations using the methods of \protect \cite{Miranda2020I} are shown in the gray circles and dashed line. Whilst the linear calculation provides a good estimate of the one-sided torque in the low mass case it fails to accurately predict total torques at high values of $\beta$. At high values of $\beta$ total torques have an additional dependence on $\Mp$ owing to the excitation of the non-linear  co-rotation (horseshoe) torque. 
    }
    \label{fig:torque_summary}
\end{figure}
%%%%%%%%%%%%%%%%%%%%%%%%%%%%%%%%%%%%%%%%%%%

%%%%%%%%%%%%%%%%%%%%%%%%%%%%%%%%%%%%%%%%%%%
\begin{figure*}
    \centering
    \includegraphics[width=\linewidth]{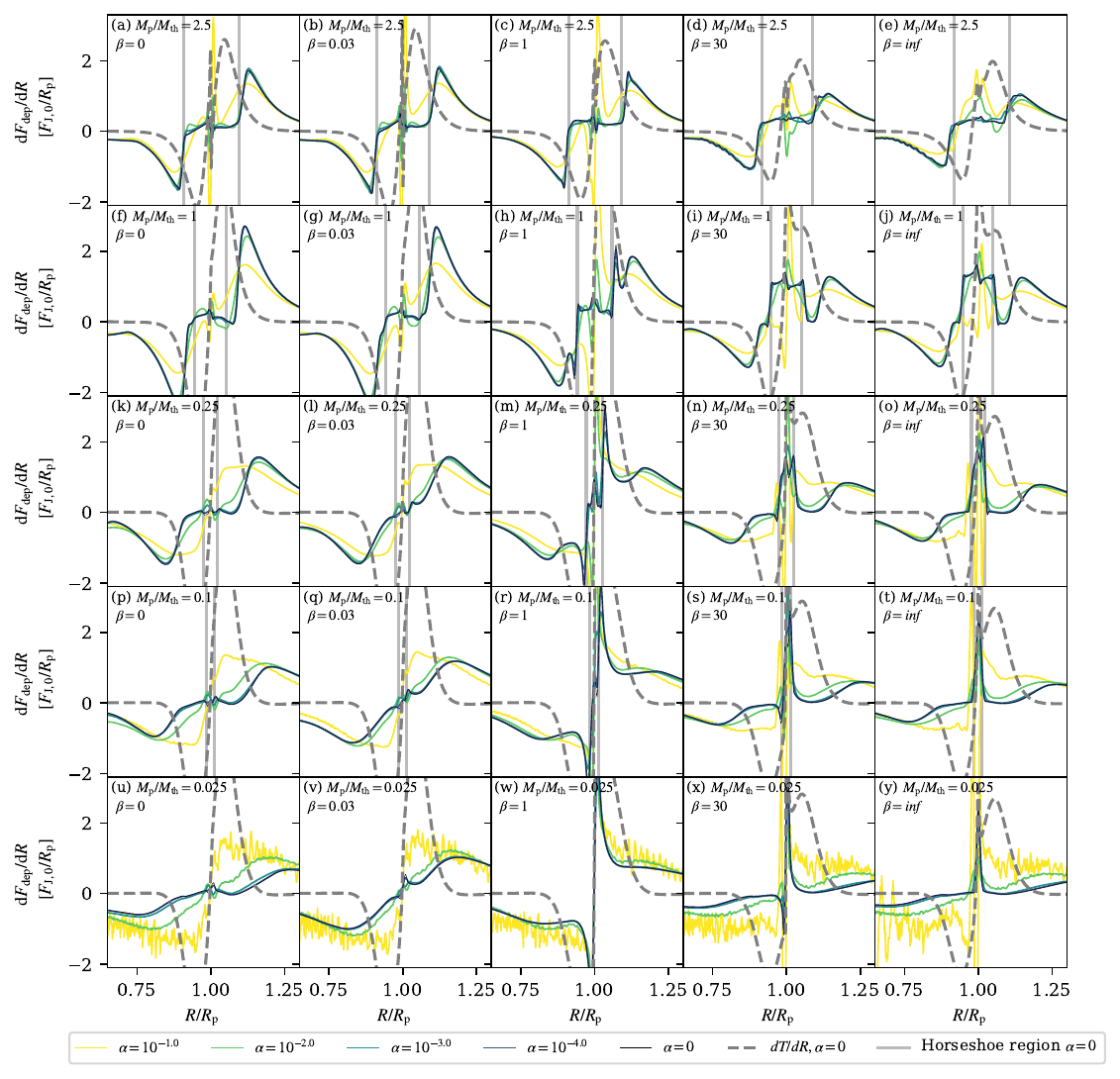}
    \caption{Angular momentum deposition $\mathrm{d}\Fdep/\mathrm{d}R$ as a function of planetary mass $\Mp$ (or $\qth$)(rows), cooling time $\beta$ (columns) and viscosity $\alpha$ (colours, see legend). The dashed gray line shows the inviscid excitation torque density $\mathrm{d}T/\mathrm{d}R$ for each $\Mp$ and $\beta$ (viscosity has a weak effect on torque excitation, see Figure \ref{fig:example_phenomology_alpha}), and the solid gray lines show the width of the horseshoe region as measured from $\alpha = 0$ simulations. Note that $\alpha =0$ and $\alpha = 10^{-4}$ cases often cannot be distinguished by eye. As mass increases shock-driven deposition (which peaks between $0.1$ and $0.2 \Rp$ away from the planet) becomes more dominant, as mass decreases linear damping mechanisms like viscosity and cooling for $\beta\sim 1$ become more important. The positive deposition structure of fixed width (often bracketed by the gray vertical lines) around $\Rp$ is caused by the co-rotation/horseshoe torque. The additional oscillations in the profile for the $\Mp=0.025 \Mth$ case at high viscosity are due to numerical issues.}
    \label{fig:fdep_grid}
\end{figure*}
%%%%%%%%%%%%%%%%%%%%%%%%%%%%%%%%%%%%%%%%%%%

%%%%%%%%%%%%%%%%%%%%%%%%%%%%%%%%%%%%%%%%%%%

\subsection{Inviscid Torque Excitation}
\label{sec:Tex}

%%%%%%%%%%%%%%%%%%%%%%%%%%%%%%%%%%%%%%%%%%%

We also briefly consider how torque excitation is impacted by varying not only $\beta$ but also $\Mp$, at fixed value of $\alpha$. For this exercise we choose $\alpha=0$, which allows us to compare our results for $\mathrm{d}T/\mathrm{d}R$ to linear calculations as outlined in Section \ref{sec:lincalc}. 

In Figure \ref{fig:torque_comparison}, we show torque excitation profiles as a function of $\Mp$ and cooling timescale and compare them to linear torque calculations (black curves). The linear calculation accurately captures the shape of the excited torque for low mass planets $\Mp\geq\Mth$ ($\qth\geq 1$), especially at lower values of $\beta$. For a planetary mass of $2.5 \Mth$ there is an overall reduction in magnitude of $\mathrm{d}T/\mathrm{d}R$ compared to other $\Mp$ due to the already significant gap opening ($\delta \Sigma/\Sigma_0 \sim 0.2$) that occurred prior to the end of the simulation ramp up period. In addition, for $\Mp\gtrsim \Mth$ ($\qth\gtrsim 1$) equation (\ref{eq:lsh}) predicts $\lsh\lesssim \Hp$, so that shock damping begins prior to the wake leaving the torque excitation region which reduces the magnitude of the excited torque.

In the adiabatic regime, $\beta=\infty$, shown in Figure \ref{fig:torque_comparison}(e), the linear calculation predicts a sharp torque structure at $\Rp$ caused by the co-rotation torque. In all of our $\beta=\infty$ simulations, even the very low mass $\Mp/\Mth = 0.025$ case, this is not actually representative of the structure of the excited horseshoe torque, which appears as a somewhat broadened positive peak. This difference is likely due to the classic linear co-rotation torque being a valid approximation only in the limit of very low planetary mass. As planetary mass increases we find that the width of this co-rotation torque excitation feature increases. 

At intermediate values of $\beta$, the structure of torque excitation close to the planet is different from $\beta=\infty$ case.  Consider the $\beta = 10$ case as shown in Figure \ref{fig:torque_comparison}(d). The linear calculation features sharp maxima and minima of finite width on either side of $\Rp$, which are slightly lower in magnitude compared to the $\beta = \infty$ case. As $\beta$ decreases again to $1$ and then to $0.1$ this feature gets smoothed out and merges with the main Lindblad torque feature before disappearing entirely. This near-$\Rp$ feature also exists in our simulation results. For all $\beta \leq 10$, the torque structure clearly approaches the linear results as $\Mp$ decreases, and there is a much better agreement between the $\Mp = 0.025 \Mth$ torque and the linear torque compared to the adiabatic case.

In Figure \ref{fig:torque_summary} we show the total ($T$) and one-sided ($T_\mathrm{OSL}$) torques defined by equations (\ref{eq:summary_T}) and (\ref{eq:summary_tosl}), as functions of $\beta$ and $\Mp$.  The linear calculations of $T$ and $T_\mathrm{OSL}$ (gray curves) are in general agreement with the results of \citet[][see their Fig. 5]{Miranda2020I}, who considered different background $\Sigma$ and $T$ profiles. We find that linear theory provides excellent estimates for the one-sided torque excited by sub-thermal mass planets ($\qth\leq 1$), see panel (a), although the torque excitation for the thermal and super-thermal mass cases ($\qth\geq 1$) $T_\mathrm{OSL}$ goes down owing to early gap opening and shock damping. The story is different for the total (migration) torque, see Figure \ref{fig:torque_summary}(b). For small values of $\beta$ linear calculations provide a good estimate for the total torque, however as $\beta$ increases and the co-rotation torque becomes active the results of linear calculations and simulations begin to diverge. This additional mass dependence is expected from studies of the co-rotation torque \citep[e.g.][]{masset_on_2009, paardekooper_torque_2010, brown_horseshoes_2024}.

%%%%%%%%%%%%%%%%%%%%%%%%%%%%%%%%%%%%%%%%%%%
%%%%%%%%%%%%%%%%%%%%%%%%%%%%%%%%%%%%%%%%%%%

\section{Parameter Variation}
\label{sec:parvar}

%%%%%%%%%%%%%%%%%%%%%%%%%%%%%%%%%%%%%%%%%%%

Having established in Section \ref{sec:results} some of the key features of $\mathrm{d}\Fdep/\mathrm{d}R$ profiles that are affected by viscosity and cooling, we now consider, in Figure \ref{fig:fdep_grid}, how the $\mathrm{d}\Fdep/\mathrm{d}R$ profiles vary also with planet mass. This allows us to get a broader view on how the three parameters $\alpha$, $\beta$ and $\qth$ jointly affect the pattern of angular momentum deposition. 

Starting with the the inviscid ($\alpha=0$), globally isothermal ($\beta=0$) case (black curves in the left column), one can see that increasing $\Mp$ concentrates angular momentum deposition nearer to the planet, as expected from the weakly non-linear theory \citep{GR01, R02} since $\lsh\propto \Mp^{-2/5}$, see equation (\ref{eq:lsh}). This is not as visually obvious when comparing the $\qth = \Mp/\Mth = 1$ and $\qth = 2.5$ isothermal cases than for lower masses, likely due to nonlinearity affecting torque excitation and noticeable gap development ($\delta \Sigma/\Sigma_0\sim 20\%$) for $\Mp\gtrsim \Mth$ at the time when angular momentum fluxes are measured. 

In the inviscid adiabatic, $\beta = \infty$, case (right column) a similar $\mathrm{d}\Fdep/\mathrm{d}R$ pattern is visible away from the  radial location of the planet, but there is also a positive contribution due to the horseshoe torque near $\Rp$. We find that the width of the horseshoe region increases with mass as expected from theoretical studies of the horseshoe torque \citep{masset_on_2009,brown_horseshoes_2024}. Also, in the $2.5 \Mth$ case, Figure \ref{fig:fdep_grid}(e), there is a dramatic reduction of $\mathrm{d}\Fdep/\mathrm{d}R$ in this region, likely owing to some non-linear effects, possibly involving shock heating. 

\begin{figure}
    \centering
    \includegraphics[width=\linewidth]{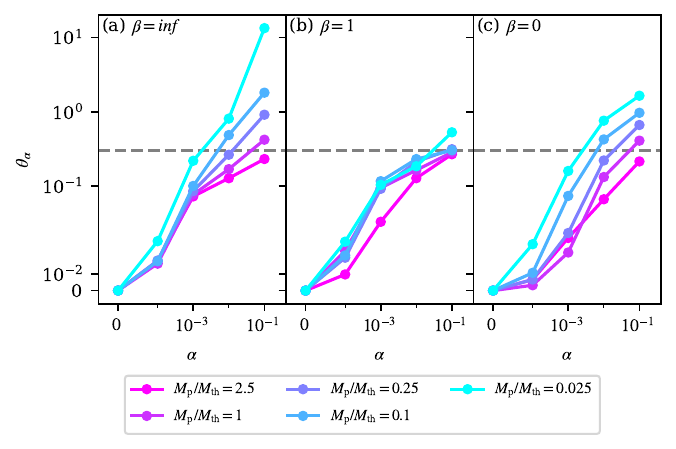}
    \caption{The normalised difference metric for viscosity $\theta_{\alpha}$ (equation \ref{eq:def_theta_alpha}) as a function of viscosity $\alpha$, planetary mass $\Mp$ (or $\qth$), and a $\beta$-cooling parameter. Line colours illustrate different masses (see legend), and each panel shows results for a different value of $\beta$ (only a small subset of $\beta$ parameters are shown). The dashed gray line is set at $\theta_{\alpha} = 0.3$, which we choose as the point above which we consider viscosity to start being important for wave damping relative to other processes (cooling and non-linearity). 
    }
    \label{fig:theta_plot}
\end{figure}

We find that cooling and viscosity become more important as the planetary mass decreases, as expected for linear effects. At the most impactful value of cooling timescale, $\beta = 1$, we find that for the super-thermal (Figure \ref{fig:fdep_grid}(c)) mass planet there is little structural difference from the isothermal case, whereas in the $\Mp = 0.1 \Mth$ case, Figure \ref{fig:fdep_grid}(r), cooling is by far the most important wave damping mechanism. A peculiar pattern of $\mathrm{d}\Fdep/\mathrm{d}R$ for $\beta\sim 1$ (described in Figure \ref{fig:example_phenomology_beta}) is also easily traced in Figure \ref{fig:fdep_grid}: a second peak (trough) caused by cooling-driven damping appears closer to the planet in the outer (inner) disc, in addition to the main $\mathrm{d}\Fdep/\mathrm{d}R$ feature caused by non-linear dissipation.  

This pattern also holds for viscosity. Excluding the the adiabatic $\Mp=2.5 \Mth$ case (Figure \ref{fig:fdep_grid}(e)) which seems to have some additional structure formation at moderate values of $\alpha$, we find that viscosity has a larger impact on $\mathrm{d}\Fdep/\mathrm{d}R$ for lower $\Mp$. In the isothermal $\Mp=2.5 \Mth$ case, Figure \ref{fig:fdep_grid}(a), even a high $\alpha = 0.1$ has only a mild effect on $\mathrm{d}\Fdep/\mathrm{d}R$ structure, while at the low-$\Mp$ end, Figure \ref{fig:fdep_grid}(p) and (u), a lower viscosity of $\alpha = 10^{-2}$ has a considerable impact on the deposition pattern. When the viscosity does considerably affect $\mathrm{d}\Fdep/\mathrm{d}R$, its main effect is to cause deposition to happen closer to the planet (compared to the inviscid case). Nonetheless, it is clear that the idea of a quasi-instantaneous, strong viscous wave damping does not hold, even in the our low $\Mp$, high $\alpha$ simulations, as at no time does the angular momentum deposition structure closely follow the excitation torque structure (dashed gray curve). 

%%%%%%%%%%%%%%%%%%%%%%%%%%%%%%%%%%%%%%%%%%%%%%%%%%

\subsection{Quantifying the differences in angular momentum deposition}
\label{sec:quant_results}

%%%%%%%%%%%%%%%%%%%%%%%%%%%%%%%%%%%%%%%%%%%%%%%%%%

There are multiple ways in which one can quantitatively assess the role of each process --- viscosity, cooling and nonlinearity--- in damping planet-induced density waves. For example, one can try to directly measure the contribution of each process to $\mathrm{d}\Fdep/\mathrm{d}R$ at every $R$ in a given simulation. In practice, however, separating the contribution due to each process is a rather tricky task, which is strongly affected by numerical noise, especially for low $\Mp$ due to the low wave amplitude. 

Instead, we have chosen a different route of defining a scalar quantitative measure of each individual process in the three-dimensional space of $\alpha$, $\beta$, and $\qth$ based on the $\mathrm{d}\Fdep/\mathrm{d}R$ structure. To do so, we define the following square difference metrics:
\begin{align}
\label{eq:def_theta_alpha}
    \theta_{\alpha}(\Mp, \beta | \alpha) &\equiv \sqrt{ \frac{\int \left( \frac{\mathrm{d} \Fdep }{\mathrm{d} R}{(\alpha)}
    - \frac{\mathrm{d} \Fdep }{\mathrm{d} R} (\alpha= 0) \right)^2
    dR }{\int \left(\frac{\mathrm{d} \Fdep }{\mathrm{d} R} (\alpha= 0) \right)^2 dR}}\,, \\
\label{eq:def_theta_beta_iso}
    \theta_{\beta, \mathrm{iso}}(\Mp, \alpha | \beta) &\equiv \sqrt{ \frac{\int \left( \frac{\mathrm{d} \Fdep }{\mathrm{d} R}{(\beta)}
    - \frac{\mathrm{d} \Fdep }{\mathrm{d} R} (\beta= 0) \right)^2
    dR }{\int \left(\frac{\mathrm{d} \Fdep }{\mathrm{d} R} (\beta= 0) \right)^2 dR}}\, \text{, and} \\
    \label{eq:def_theta_beta_adi}
    \theta_{\beta, \mathrm{adi}}(\Mp, \alpha|\beta) &\equiv \sqrt{ \frac{\int \left( \frac{\mathrm{d} \Fdep }{\mathrm{d} R}{(\beta)}
    - \frac{\mathrm{d} \Fdep }{\mathrm{d} R} (\beta= \infty) \right)^2
    dR }{\int \left(\frac{\mathrm{d} \Fdep }{\mathrm{d} R} (\beta= \infty) \right)^2 dR}}\,,
\end{align}
where $\mathrm{d} \Fdep/\mathrm{d} R$ is calculated from a simulation with the given $\alpha, \beta, \Mp$ (or $\qth$) and each metric explores the variation of the last parameter in its definition. These metrics describe how the variation of a particular parameter affects the overall shape of $\mathrm{d}\Fdep/\mathrm{d}R$ (in the integrated sense) relative to the $\mathrm{d}\Fdep/\mathrm{d}R$ curve obtained for some reference value of this parameter. It is important to remember that in general all three processes --- viscosity, cooling and shock dissipation --- operate and affect wave damping in a given simulation, but $\theta$-metrics are sensitive to the variation of only one of them, keeping parameter values characterizing other processes unchanged.   

For example, $\theta_{\alpha}(\Mp, \beta | \alpha)$ measures the differences in $\mathrm{d}\Fdep/\mathrm{d}R$ arising as we vary $\alpha$ away from the inviscid case of $\alpha=0$, which we have chosen (quite naturally) to be the reference state. When exploring the variation of $\beta$, the choice of the reference state is not so obvious. Both $\beta=0$ (isothermal) and $\beta=\infty$ (adiabatic) are natural limits to which an arbitrary $\beta$ state can be compared. Firstly, there is no singular obvious choice that we can define to be the background state in absence of all thermodynamic effects. At each extremum for our choice of background temperature structure (globally isothermal) angular momentum deposition happens solely due to shock dissipation and viscosity, and either the $\beta=0$ or $\beta=\infty$ approximations may be valid in different regions of protoplanetary discs. Secondly, angular momentum deposition is often more similar between simulations with either an isothermal or adiabatic equation of state than they are with a simulation with an intermediate value of $\beta$.

For these reasons, instead of choosing a single point of comparison, we opted to define two $\theta$-metrics characterizing $\beta$ variation: $\theta_{\beta, \mathrm{iso}}$ defined by equation (\ref{eq:def_theta_beta_iso}) compares $\mathrm{d}\Fdep/\mathrm{d}R$ relative to the isothermal case, while $\theta_{\beta, \mathrm{adi}}$ defined by equation (\ref{eq:def_theta_beta_adi}) compares it to the adiabatic case. This selection will allow us to identify the separate regions of parameter space where either the adiabatic or isothermal approximations for angular momentum deposition are valid.

Note that we are not providing any metric for the variation of $\Mp$ (or $\qth$) as the effects of wave non-linearity on its damping have been explored in many studies in the past \citep{GR01,Dong2011b,Cimerman2021}. In our present work the effects of varying $\Mp$ are obvious in the behaviour of $\theta$-metrics, see Section \ref{sec:critical_values}.

%%%%%%%%%%%%%%%%%%%%%%%%%%%%%%%%%%%%%%%%%%%%%%%%%%
\begin{figure}
    \centering
    \includegraphics[width=\linewidth]{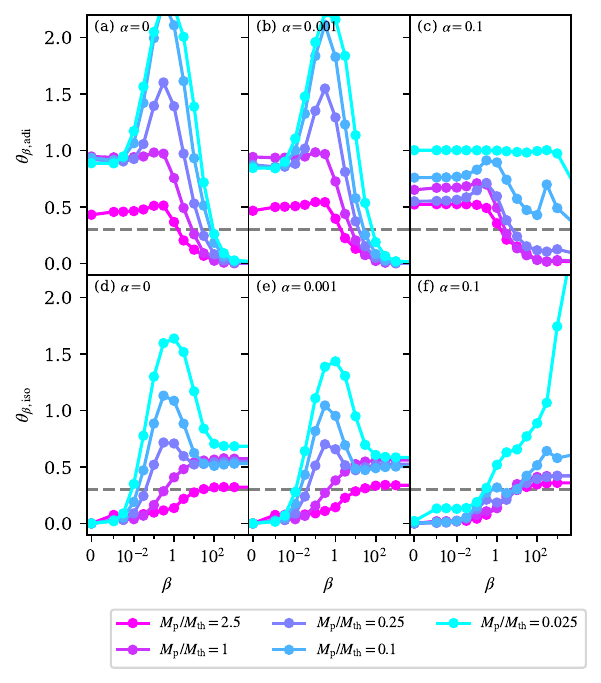}
    \caption{The normalised difference metric for cooling time-scales: (a, b, c) show the difference of each simulation to the adiabatic case, $\theta_{\beta, \mathrm{adi}}$ (Equation \ref{eq:def_theta_beta_adi}) and (d, e, f) show the difference metric compared to isothermal simulations, $\theta_{\beta, \mathrm{iso}}$ (Equation \ref{eq:def_theta_beta_iso}). Each column shows the results for a different value of $\alpha$, and line colours refer to different planetary masses (see legend). 
    The dashed gray line is set at $0.3$, and we consider any simulation where $\theta_{\beta, \mathrm{adi}}$ ($\theta_{\beta, \mathrm{iso}}$) is above this value to have thermal effects on wave damping significantly different from those in an adiabatic (isothermal) disc.
    }
    \label{fig:theta_beta_plot}
\end{figure}
%%%%%%%%%%%%%%%%%%%%%%%%%%%%%%%%%%%%%%%%%%%%%%%%%%

The integration range in the definitions (\ref{eq:def_theta_alpha})-(\ref{eq:def_theta_beta_adi}) is set between $0.5\,\Rp$ and $2\,\Rp$. We chose to exclude the inner and outer parts of the simulation domain as we found that in the adiabatic and highly viscous case ($\beta=\infty,\alpha=0.1$) there are additional viscous effects related to our boundary conditions. Moreover, we also remove the horseshoe region from the integration range, the inner and outer radial boundaries of which are shown in gray lines on Figure \ref{fig:fdep_grid}. We choose to ignore this region as we are interested only in damping of the propagating planet-driven density waves capable of transporting their angular momentum far from their excitation site. Also, angular momentum deposition in the horseshoe region is highly dependent on the background disc temperature gradient, and in any case the co-orbital contribution to $\mathrm{d}\Fdep/\mathrm{d}R$ typically disappears after a few hundred orbits in an inviscid disc. A well-defined horseshoe region was not detected in our simulations of $0.025 \Mth$ planets and, as such, we do not exclude any co-orbital range from the calculation in that case. As with the plots shown in Figures \ref{fig:example_phenomology_alpha}, \ref{fig:example_phenomology_beta} and \ref{fig:fdep_grid}, we smooth out the extracted $\mathrm{d} \Fdep/\mathrm{d}R$ profiles using a Savitzky-Golay filter before using them to calculate our metrics. 

We show the results for the normalised difference metric in viscosity, $\theta_{\alpha}$, in Figure \ref{fig:theta_plot}. The structure of $\theta_{\alpha}(\alpha)$ is fairly simple to understand. It grows monotonically with $\alpha$, reflecting the ever-increasing importance of viscosity as $\alpha$ increases. This growth is generally faster as $\Mp$ (or $\qth$) decreases, simply because the otherwise dominant non-linear wave damping is less important in the low-$\Mp$ regime, so that viscosity can provide a relatively larger contribution to $\mathrm{d}\Fdep/\mathrm{d}R$. This trend is less obvious for $\beta=1$ as shown in Figure \ref{fig:theta_plot}(b), since in this case linear cooling plays a very important role in wave damping, as we will see next. We also find that viscosity is generally somewhat more effective at modifying wave damping in the adiabatic rather than isothermal regime --- likely due to the weaker torque excitation in this case making disc-planet coupling less non-linear. 

%%%%%%%%%%%%%%%%%%%%%%%%%%%%%%%%%%%%%%%%%%%%%%%%%%
\begin{figure*}
    \centering
    \includegraphics[width=\linewidth]{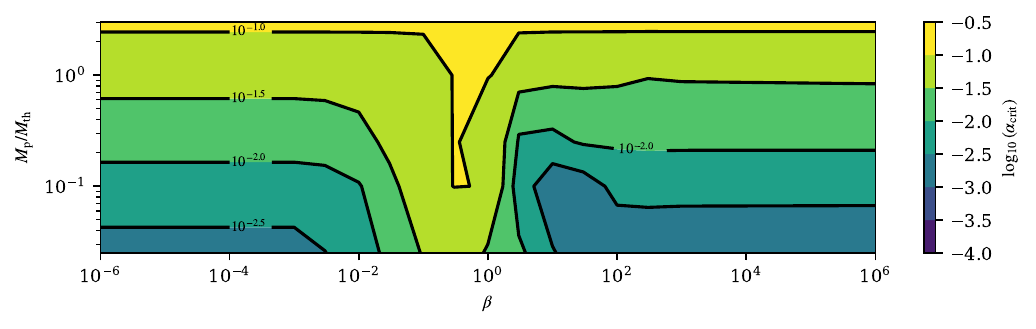}
    \caption{
    Contour map of the critical value of $\alpha$ below which viscosity can be considered as an unimportant factor in wave damping, as a function of $\Mp/\Mth$ and $\beta$. We define the critical value of $\alpha$ based on the condition $\theta_{\alpha}(\alpha) = \theta_\mathrm{crit}$, with $\theta_\alpha$ and $\theta_\mathrm{crit}$ given by equations (\ref{eq:def_theta_alpha}) and  (\ref{eq:theta_crit}). See text for details.}
    \label{fig:critical_alpha}
\end{figure*}
%%%%%%%%%%%%%%%%%%%%%%%%%%%%%%%%%%%%%%%%%%%%%%%%%%

In Figure \ref{fig:theta_beta_plot}, we show the normalised difference metrics for cooling, $\theta_{\beta, \mathrm{adi}}$ and $\theta_{\beta, \mathrm{iso}}$. Typically, for $\alpha\leq 0.01$ these metrics peak near $\beta = 1$, before going down towards the adiabatic or isothermal limits. This clearly indicates that there is no gradual, monotonic transition between $\beta=0$ and $\beta=\infty$ regimes, and instead cooling at $\beta\sim 1$ is dramatically different from either of these limits and is very effective at wave damping \citep{Miranda2020I,Miranda2020II}. In addition, we find that $\theta_{\beta}$ is often lower for higher mass planets, with cooling being most significant in the low mass case. In fact, for $2.5 \Mth$ planet simulations, we do not see a very pronounced maximum, indicating that the changes of angular momentum deposition induced by linear cooling are minimal compared to wave damping due to shock formation. It is possible in this high $\Mp$ case that much of the variation of $\theta_\beta$ with $\beta$ is simply due to changes in overall torque excitation, see Figure \ref{fig:torque_summary}. Referring back to Figure \ref{fig:fdep_grid}(a)-(e) this seems to be the case as we do not see the additional peak of linearly driven angular momentum deposition that we see in the lower mass cases (e.g. compare with $\Mp=0.25\Mth$ case).

At high viscosity, for $\alpha = 0.1$, the data tell a somewhat different story. Consider $\theta_{\beta, \mathrm{iso}}$ shown in Figure \ref{fig:theta_beta_plot}(f). Excluding the lowest mass case -- which has some additional numerical issues -- we see that  $\theta_{\beta, \mathrm{iso}}$ is typically lower than it would be in the inviscid case, and unlike the inviscid case does not exhibit a clear maximum near $\beta \approx 1$; instead, it climbs near monotonically to a single lower value at the adiabatic extreme. In this case, linear damping from viscosity must be more efficient than damping due to cooling and dominate as a damping source. The effects of cooling are not completely removed, though some of the variation of $\theta_{\beta, \mathrm{iso}}$ with $\beta$ may in fact be due to the $\beta$-dependence of the excited torque.

%%%%%%%%%%%%%%%%%%%%%%%%%%%%%%%%%%%%%%%%%%%%%%%%%%

\subsection{Threshold level for $\theta$-metrics}
\label{sec:thresh}

%%%%%%%%%%%%%%%%%%%%%%%%%%%%%%%%%%%%%%%%%%%%%%%%%%

The $\theta$-metrics are designed to be zero when the parameter that they correspond to is set to its reference value, e.g. $\alpha=0$ for $\theta_{\alpha}$. The metric grows as this parameter starts to deviate from its reference value, indicating the increasing role of the corresponding physical process in wave damping. It is natural to try to come up with a single critical (or threshold) value of $\theta$, beyond which a corresponding process would be considered as playing an important role in wave damping (when compared to the reference parameter choice). Establishing such a critical value $\theta_{\mathrm{crit}}$ would allow us to characterize the importance of a given wave-damping process across the full three-dimensional space of $\alpha,\beta,\Mp$ in a uniform fashion, which we do next in Section \ref{sec:critical_values}. 

By visually inspecting the characteristic shapes of $\mathrm{d}\Fdep/\mathrm{d}R$ curves in Figure \ref{fig:fdep_grid} and their evolution as the various parameters are varied, we came up with  the following critical value of all three $\theta$-metrics, 
\begin{equation}
    \theta_{\mathrm{crit}} = 0.3.
    \label{eq:theta_crit}
\end{equation}
This choice is necessarily somewhat subjective, but it should not affect the qualitative conclusions of this work. 

We show $\theta_{\mathrm{crit}}$ in Figures \ref{fig:theta_plot} and \ref{fig:theta_beta_plot} as a dashed horizontal line. We illustrate the use of $\theta_{\mathrm{crit}}$ concept by examining Figure \ref{fig:theta_plot}(a), which represents an adiabatic case, $\beta=\infty$. One can see that for $\Mp=0.025\Mth$ ($\qth=0.025$) the criterion $\theta_\alpha>\theta_{\mathrm{crit}}$ predicts that viscosity strongly modifies wave damping for $\alpha\gtrsim 10^{-3}$, whereas for $\Mp=1\,\Mth$ ($\qth=1$) the same is true only for $\alpha\gtrsim 0.03$. This logic is developed further in the next Section.

%%%%%%%%%%%%%%%%%%%%%%%%%%%%%%%%%%%%%%%%%%%%%%%%%%

\subsection{Critical values of \protect $\alpha$ and \protect $\beta$}
\label{sec:critical_values}

%%%%%%%%%%%%%%%%%%%%%%%%%%%%%%%%%%%%%%%%%%%%%%%%%%

We now use the concept of $\theta_{\mathrm{crit}}$ to illustrate the importance of each wave damping process using two-dimensional representation in a space of two parameters. Starting with viscosity, 
we define the critical viscosity $\alpha_{\mathrm{crit}}(\beta, \Mp)$ as the lowest value of $\alpha$ for which $\theta_{\alpha} > \theta_{\mathrm{crit}}$, as a function of $\beta$ and $\Mp$. Based on our logic, $\alpha_{\mathrm{crit}}(\beta, \Mp)$ should be interpreted as the value of $\alpha$ below which viscosity is unimportant for angular momentum deposition given the values of $\beta$ and $\Mp$. 
We estimate $\alpha_{\mathrm{crit}}(\beta, \Mp)$ using two point interpolation over $\theta_{\alpha}(\alpha)$ curves in log-log space, see Figure \ref{fig:theta_plot}. In cases where no values of $\alpha$ lead to $\theta_{\alpha} > \theta_{\mathrm{crit}}$ (i.e. viscosity is always unimportant for wave damping) within our grid of $\beta$ and $\Mp$, we instead report the lower limit of $\alpha_{\mathrm{crit}}$ as $10^{-1}$ in our plots. A contour map of $\alpha_{\mathrm{crit}}$ is shown in Figure \ref{fig:critical_alpha} (a plot showing the specific values of $\alpha_{\mathrm{crit}}$ used to generate the contour map can be found in Appendix \ref{sec:additional_plots}). 

%%%%%%%%%%%%%%%%%%%%%%%%%%%%%%%%%%%%%%%%%%%%%%%%%%
\begin{figure*}
    \centering
    \includegraphics[width=\linewidth]{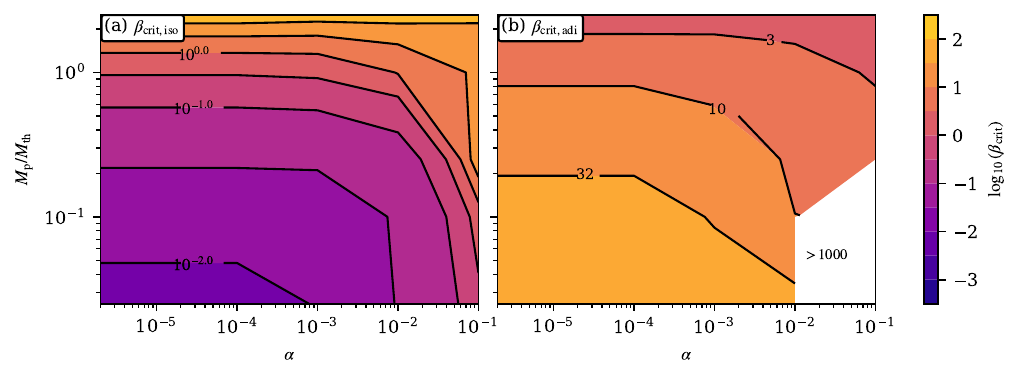}
    \caption{
    Contour map of the critical values of $\beta$ as a function of planetary mass and viscosity: (a) the critical values of $\beta$ above which the dynamics of wave damping cannot be accurately approximated as isothermal, (b) the critical values of $\beta$ below which wave damping cannot be accurately approximated as adiabatic. The critical $\beta$ values correspond to $\theta_{\beta, \mathrm{iso}}$ or $\theta_{\beta, \mathrm{adi}}$ (defined by equations (\ref{eq:def_theta_beta_iso}) and (\ref{eq:def_theta_beta_adi})) first reaching $\theta_\mathrm{crit}$ starting from $\beta = 0$ or $\beta = \infty$ for the isothermal and adiabatic cases, respectively. See text for details.}
    \label{fig:critical_beta}
\end{figure*}
%%%%%%%%%%%%%%%%%%%%%%%%%%%%%%%%%%%%%%%%%%%%%%%%%%

This figure shows that, given our adopted ranges for $\beta$ and $\Mp$, the smallest value of $\alpha_{\mathrm{crit}}$ is $10^{-2.9}$, indicating that below this value angular momentum deposition due to viscosity can be ignored. But in the vast majority of our $(\beta,\Mp)$ parameter space viscosity described by much higher values of $\alpha$ can be ignored. For example, for planets with $\Mp>1\, \Mth$ we find that $\alpha_{\mathrm{crit}}$ is always larger than $10^{-1.5}$. In all regimes $\alpha_{\mathrm{crit}}$ increases monotonically with mass indicating that, as expected, the increased importance of non-linear wave damping makes viscosity less important in comparison. At intermediate values of $\beta\sim 1$, $\alpha_{\mathrm{crit}}$ is large even for much lower $\Mp$, indicating that viscosity is also not important for wave damping when cooling is very efficient.

Moving on to the role of cooling, we define $\beta_{\mathrm{crit, iso}}$ as the \emph{largest} value of $\beta$ for which $\theta_{\beta, \mathrm{iso}} \leq \theta_{\mathrm{crit}}$. Similarly, $\beta_{\mathrm{crit, adi}}$ is defined as the \emph{smallest} value of $\beta$ for which $\theta_{\beta, \mathrm{adi}} \leq \theta_{\mathrm{crit}}$. We calculate $\beta_{\mathrm{crit, iso}}$ and  $\beta_{\mathrm{crit, adi}}$ via logarithmic two-point interpolation of $\beta$ as applied to the values of $\theta_{\beta}$ adjacent to $\theta_{\mathrm{crit}}$. In the cases where the only value of $\theta_{\beta, \mathrm{adi}}$ that was less than $\theta_{\mathrm{crit}}$ was $\beta = \infty,$ we instead provide an upper limit. 

Contour maps of $\beta_{\mathrm{crit}}$ are shown in Figure \ref{fig:critical_beta} (a plot showing the specific values of $\beta_{\mathrm{crit}}$ used to generate the contour maps is included in Appendix \ref{sec:additional_plots}). Let us first focus on $\beta_{\mathrm{crit, iso}}$, as shown in Figure \ref{fig:critical_beta}(a). We find that the range of $\beta$ for which the angular momentum deposition is well represented by the isothermal limit increases with both planetary mass and viscosity. In the inviscid and low-mass limit, $\beta < 0.01$ is required for the isothermal approximation to be valid for wave damping, whereas in the high-mass inviscid limit the isothermal approximation is reasonable for values of $\beta$ as large as $\sim20$. In the low-mass, high-$\alpha$ limit this approximation remains reasonable up to $\beta \sim 0.25$. These variations can be naturally traced back to the varying roles of non-linearity and viscosity in wave damping.

Similarly in Figure \ref{fig:critical_beta}(b), we find that the range of values for which wave damping can be well described by the adiabatic approximation increases with higher masses but only weakly depends on viscosity --- in this case as our limit is $\beta = \infty$, this conclusion follows from a decrease of $\beta_{\mathrm{crit, adi}}$. For high viscosities, $\alpha = 0.1$ and low masses $\Mp/\Mth \leq 0.1$ we found no values of $\beta \neq \infty$ for which $\theta_{\beta, \mathrm{adi}} < \theta_{\mathrm{crit}}$. This appears to be a numerical effect connected to the oscillations seen in the $\mathrm{d} \Fdep/\mathrm{d}R$ profiles for this $\alpha$, see Figure \ref{fig:fdep_grid}(x) and (y). Given the shape of the plot outside of this region, we suggest that for practical applications the wave damping in this region can be approximated as near adiabatic for $\beta \gtrsim 40$.

In Appendix \ref{sec:sec_order}, we calculate critical values from simulations using the more traditional second-order smoothed (Plummer) potential. We find no significant differences between the maps produced by this calculation, indicating that the boundaries of regions dominated by a particular damping process are rather insensitive to the choice of smoothed potential, given a correctly estimated smoothing length. A link to a Zenodo repository containing all the $\Fdep$ profiles used in this analysis is contained in the data availability section, and these can be used freely for more specific analysis.

%%%%%%%%%%%%%%%%%%%%%%%%%%%%%%%%%%%%%%%%%%%%%%%%%%
%%%%%%%%%%%%%%%%%%%%%%%%%%%%%%%%%%%%%%%%%%%%%%%%%%

\section{Discussion}
\label{sec:discussion}

%%%%%%%%%%%%%%%%%%%%%%%%%%%%%%%%%%%%%%%%%%%%%%%%%%

This study has explored in a uniform fashion the relative contributions provided by $\alpha$-viscosity, $\beta$-cooling and non-linear wave evolution to density wave excitation by planets and wave damping. By considering a wide range of parameters and using 2D disc simulations we were able to identify the regions of the parameter space where each of the processes dominates wave damping. 

Based on this work we can conclude that wave damping due to viscosity should be unimportant in most protoplanetary discs. Indeed, observational constraints suggest viscosities of $\alpha \lesssim 10^{-3}$ to be typical in these discs \citep{Flaherty2015,Flaherty2017,Rafikov2017,rosotti_empirical_2023}, and our theoretical modelling shows that viscosity is an unimportant factor in wave damping for $\alpha \leq 10^{-2.9}$ even for very low mass planets. For thermal and super-thermal mass planets ($\qth\gtrsim 1$), we find that viscosity is unimportant even for larger values of $\alpha \leq 10^{-1.5}$, regardless of our thermodynamic assumptions. In discs with cooling time-scales $\beta\sim 1$, viscosity with $\alpha \leq 10^{-1.5}$ is unimportant for wave damping even for lower mass planets, down to $\Mp=0.025\Mth$. 

It is important to keep in mind that these statements are true only regarding the role of viscosity in damping density waves, but not for forming substructures (rings and gaps) in protoplanetary discs. Indeed, planet-driven structure formation is affected by viscosity in two ways: firstly, through its contribution to wave damping, and secondly, through its tendency to erase any inhomogeneities in density distribution in discs. The latter process is very effective and has been shown by \citet[][see their Figures 8(e),(k), 9(e),(k)]{Miranda2020II} to strongly suppress structure formation by planet-driven density waves even for values of $\alpha\sim$, which we found to cause negligible wave damping in this work. This double manifestation of viscous effects should always be kept in mind when thinking about gap formation in protoplanetary discs. 

Cooling, however, is likely to be quite important for density wave damping in the outer parts (at tens of AU) of many protoplanetary discs, as the expected values of $\beta$ are often found to be around 1 in this region \citep[e.g.][]{Ziampras2023}. At viscosity levels relevant to protoplanetary discs, the effects of cooling on wave evolution converges to the isothermal limit for $\beta < 0.01$ and to adiabatic limit when $\beta > 80$. For $\beta$ values in between the effect of cooling on wave damping is very strong and must be modelled explicitly \citep{Miranda2020I,Miranda2020II}.

In this paper, in order to allow ourselves to explore a broad range of the three-dimensional parameter space we have simulated planet-disc interaction only on rather short timescale, making sure the physical convergence is reached. Based on previous studies \citep{Cimerman2021, Cordwell2024} we expect our quantitative results for $\mathrm{d} \Fdep/\mathrm{d}R$ profiles to hold generally while gaps remain shallow. Once gaps grow deep enough, there can be significant changes to both torque excitation and angular momentum deposition patterns \citep{petrovich_disk-satellite_2012, Cimerman2021}. However, our key results in Section \ref{sec:critical_values} on the role of different wave damping mechanisms as a function of $\alpha,\beta,\Mp$ should not be affected by gap development. We have also made a number of thermodynamic simplifications: simulated only discs with a constant background temperature profile, considered a simplified model of thermodynamic response (linear $\beta$-cooling) and have not considered spatial variation of the cooling function (i.e. radially-dependent $\beta$). These restrictions, along with the 2D nature of our work, can be easily relaxed in future work.

Results of our study can be used to understand wave propagation in other types of astrophysical discs. As our explored parameter space stretches into high levels of viscosity, we can apply our understanding of the roles played by different wave damping processes to wave propagation in discs of cataclysmic variables and X-ray binaries \citep{Ju2016,van2023}, acoustic waves in the boundary layers of accretion discs \citep{Belyaev2012,Belyaev2013,Coleman2022a,Coleman2022b,Dittmann2024}, or density waves driven by extreme-mass-ratio inspirals in AGN discs \citep{Dyson2026}.

%%%%%%%%%%%%%%%%%%%%%%%%%%%%%%%%%%%%%%%%%%%%%%%%%%

\subsection{Comparison to previous work}

%%%%%%%%%%%%%%%%%%%%%%%%%%%%%%%%%%%%%%%%%%%%%%%%%%

The most direct comparison to our work is provided by \cite{Miranda2020II} who also ran a suite of planet-disc interaction simulations with varied planetary mass, viscosity and cooling. Their work used a slightly more complex and physically motivated cooling function with $\beta$ as a function of radius, but they did not carry out as extensive parameter space exploration as we do here. In addition, they used a different metric to study the effects of $\alpha$, $\beta$ and $\Mp$ on wave propagation, namely $\Fwave$ (referred to as AMF in their work), unlike our present work which utilizes $\mathrm{d} \Fdep/\mathrm{d}R$ for that purpose. \citet[][see their Figure 12]{Miranda2020II} compared AMF profiles derived from simulations with a $\Mp = 0.3 \Mth$ planet, different cooling assumptions, and a variety of $\alpha$ values, finding that in most cases viscosity is unimportant for density wave damping. They also found that, depending on the planetary location in the disc (which sets the effective value of $\beta$), non-linear effects and cooling can provide comparable contribution to density wave evolution (see their Figure 7). These conclusions are fully consistent with our findings. Our present work surpasses \cite{Miranda2020II} in providing quantitative characterization of damping effects of different physical processes as well as in the extent of the parameter space explored. 

Prior analytic work on mechanisms of density wave damping has primarily focused on each physical process in isolation. In particular, \citet{Takeuchi1996} explored wave damping by viscosity only, \citet{Cassen1996}, \citet{Miranda2020I} focused on damping by radiative cooling, while \citet{GR01}, \citet{R02} explored non-linear wave damping acting in isolation \citep[although][also discussed wave damping by cooling]{GR01}. These studies have often tried to come up with a characteristic damping length for the primary azimuthal mode of the linearized wave excitation problem and did not consider several physical mechanisms operating simultaneously. However, the angular momentum deposition profiles we observe in our work are non-trivial (see Figure \ref{fig:fdep_grid}), cannot be described as a simple exponential damping, and naturally account for the full spectrum of excited azimuthal modes. 
%It is possible to show that our simulation results are congruent with a comparison of length scales from analytic theory, although as the length scale for half the wave to be dissipated by shock formation, $2 \lsh$, \citep{GR01} is not necessarily equivalent to a damping length scale we do not perform this analysis in depth. 
Our study has extended upon those previous studies by allowing for direct comparisons between combinations of different damping mechanisms and by quantifying those relationships numerically.

%%%%%%%%%%%%%%%%%%%%%%%%%%%%%%%%%%%%%%%%%%%%%%%%%%

\subsection{Implications for disc observations}
\label{sec:obs}

%%%%%%%%%%%%%%%%%%%%%%%%%%%%%%%%%%%%%%%%%%%%%%%%%%

As mentioned earlier, our study has focused on angular momentum deposition due to damping of planet-driven waves rather than on the natural outcome of that deposition --- development of gaps and other substructures in discs. However, since $\mathrm{d}\Fdep/\mathrm{d}R$ determines both the early gap evolution \citep{Cordwell2024} and the gap shape in the viscous steady state (see Appendix \ref{sec:vss}), we can comment on how different wave damping processes are likely to affect gap development and its observational prospects. These considerations may complement the results on longer term gap evolution reported in \cite{Miranda2020I,Miranda2020II}.

In inviscid discs and for $\beta\to 0$ or $\beta\to \infty$, the angular momentum deposition pattern due to non-linear wave dissipation has a single pronounced extremum on each side of $\Rp$, with no deposition happening in the radial range $(\Rp-\lsh,\Rp+\lsh)$. As shown in \citet{R02b} and \citet{Cordwell2024}, this inevitably leads to a characteristic `double gap' structure forming around the planetary orbit, at least when the gap is not too deep (relative depth of several tens of per cent). This picture would change if the deposition pattern shifts closer to $\Rp$, as we found to be the case for high viscosity and/or cooling with $\beta\sim 1$, see Section \ref{sec:parvar}, and a single gap would appear in this case \citep{Miranda2020I,Miranda2020II}. Thus, finding a double gap structure in observations may signal that, firstly, the viscosity is low, $\alpha\lesssim 10^{-2}$, and secondly, that $\beta$ is either above $\beta_{\mathrm{crit, adi}}$ (quasi-adiabatic regime) or below $\beta_{\mathrm{crit, iso}}$ (quasi-isothermal regime). The latter thermodynamic constraint could provide an independent check of the more traditional modelling of cooling times such as in \cite{Ziampras2020}. 

The former viscous constraint may in fact be superseded by a different constraint based on the diffusive effect of viscosity on the forming gap structure. As demonstrated in \cite{Cordwell2024}, viscosity starts modifying the double gap structure shifting it towards a single gap  after $\sim\Pp/(6 \pi \alpha)$ after the introduction of the planetary gravitational potential, where $\Pp$ is the planet's period. In real discs with observed double gaps this consideration may provide a more stringent constraint on $\alpha$ than the constraint based on inefficiency of viscous damping of planet-induced density waves.

The changes in $\mathrm{d} \Fdep/\mathrm{d}R$ due to cooling should also have an impact on gap structure over long viscous time scales. In Appendix \ref{sec:vss}, we show that in viscous steady state the gap structure is directly proportional to $\Fdep$ defined by equation (\ref{eq:global_vss}), and therefore we can directly link changes in $\mathrm{d} \Fdep/\mathrm{d}R$ to gap structures. As in discs with intermediate cooling timescales ($\beta \sim 1$) angular momentum deposition is more concentrated towards the planetary location, gaps formed in these discs will be narrower than those formed in equivalent isothermal discs. It is possible that this could explain why in some discs with directly imaged planets gaps are thinner than would be expected based on their mass estimated from other means, as is the case for WISPIT 2b \citep{capelleveen_wide_2025, close_wide_2025, facchini_2_2026}. A further study involving dust dynamics would be required to assess if $\mathrm{d} \Fdep/\mathrm{d}R$ changes due to cooling and viscosity could be used to provide observational constraints.

%%%%%%%%%%%%%%%%%%%%%%%%%%%%%%%%%%%%%%%%%%%%%%%%%%
%%%%%%%%%%%%%%%%%%%%%%%%%%%%%%%%%%%%%%%%%%%%%%%%%%

\section{Summary}
\label{sec:summary}

%%%%%%%%%%%%%%%%%%%%%%%%%%%%%%%%%%%%%%%%%%%%%%%%%%

In this work we have performed a systematic study of the roles played by different dissipative mechanisms --- viscosity, cooling and non-linear evolution --- in damping density waves launched by planets embedded in protoplanetary discs. We did this by analysing spatial patterns of angular momentum deposition by the density wave across the range of parameters --- $\alpha,\beta,\Mp$ (or $\qth$) --- characterizing the problem. Our key findings are summarized below.

\begin{itemize}
    \item Generally, angular momentum deposition due to non-linear wave evolution and shock formation is the most important cause of angular momentum deposition, except at very low planetary masses.
    
    \item Angular momentum deposition caused by viscous dissipation is unimportant for $\alpha \lesssim 10^{-2.9}$, even for waves driven by highly sub-thermal mass planets (down to $\Mp=0.025\Mth$). For planets with $\Mp \gtrsim \Mth$ ($\qth>1$), more easily detectable in observations, viscosity is unimportant for $\alpha \lesssim 10^{-1.5}$.

    \item The importance of angular momentum deposition due to $\beta$-cooling peaks at $\beta$ in the range $0.1-10$. For $\beta$ in this range cooling is competitive with non-linear wave damping (except in the super-thermal mass regime, $\qth>1$) and viscosity (except for highest $\alpha\gtrsim 0.1$).
    
    \item Figures \ref{fig:critical_alpha} and \ref{fig:critical_beta}, based on our simulation results, allow one to partition the space of parameters $\alpha,\beta,\Mp$ into regions with wave damping dominated by particular dissipative processes.

    \item Even for low mass planets and at high viscosity ($\alpha = 0.1$) the approximation of immediate damping of the planet-driven density waves is never valid. 
    
    \item Unambiguous detection of a `double gap' morphology of a planetary gap in observations may indicate that it formed in a region of a disc with low viscosity and either near-isothermal or near-adiabatic thermodynamics (i.e. $\beta$ not close to 1).
    
\end{itemize}

We have made simulation outputs for $\mathrm{d} \Fdep/\mathrm{d}R$ available to the community, and we hope that they can be used in future population-level models of planet-disc interactions. They can also be used as inputs for building a semi-analytical model of angular momentum deposition by planet-driven density waves that would account for both linear and non-linear dissipative processes.

\section*{Acknowledgements}

We thank Joshua Brown for useful discussions regarding the horseshoe torque. A.J.C. is funded by the Royal Society of New Zealand Te Apārangi and the Cambridge Trust through the Cambridge-Rutherford Memorial Scholarship. R.R.R. acknowledges financial support through the STFC grant ST/X001113/1 and the IAS.

\textit{Software:} \textsc{NumPy} \citep{harris_array_2020}, \textsc{SciPy} \citep{virtanen_scipy_2020}, \textsc{Matplotlib} \citep{thomas_a_caswell_matplotlibmatplotlib_2023} and \textsc{Athena++} \citep{stone_athena_2020}.

%%%%%%%%%%%%%%%%%%%%%%%%%%%%%%%%%%%%%%%%%%%%%%%%%%
\section*{Data Availability}

A custom version of \textsc{Athena++} with the Bessel-type smoothed potential as used for this work is available on GitHub: \url{github.com/cordwella/athena-disk-planet}.
$\mathrm{d}\Fdep / \mathrm{d} R$ curves and other summary results from our simulations are available on Zenodo: \url{https://doi.org/10.5281/zenodo.19369572}.
Additional simulation data will be made available upon reasonable request.



\bibliographystyle{mnras}
\bibliography{damping} % if your bibtex file is called example.bib

%\begin{thebibliography}{99}
%\end{thebibliography}

%%%%%%%%%%%%%%%%%%%%%%%%%%%%%%%%%%%%%%%%%%%%%%%%%%

%%%%%%%%%%%%%%%%% APPENDICES %%%%%%%%%%%%%%%%%%%%%

\appendix

%%%%%%%%%%%%%%%%%%%%%%%%%%%%%%%%%%%%%%%%%%%%%%%%%%
%%%%%%%%%%%%%%%%%%%%%%%%%%%%%%%%%%%%%%%%%%%%%%%%%%

\section{Solution of the viscous steady state in terms of \protect $\mathrm{d} \Fdep/\mathrm{d}R$}
\label{sec:vss}

%%%%%%%%%%%%%%%%%%%%%%%%%%%%%%%%%%%%%%%%%%%%%%%%%%

A key metric we use to judge the importance of various wave damping processes is the angular momentum deposition function $\mathrm{d} \Fdep/\mathrm{d}R$. It is employed here because it directly determines the outcome of the gap evolution process, and \cite{Cordwell2024} has shown in detail how $\mathrm{d} \Fdep/\mathrm{d}R$ can be used to accurately predict the early evolution of a gap when it is not too deep. We now also show explicitly that the viscous steady state of a gap \citep[as investigated in detail by e.g.][]{kanagawa_formation_2015, duffell_empirically_2020, dempsey_pileups_2020} can also be expressed directly in terms of $\mathrm{d} \Fdep/\mathrm{d}R$.

The one-dimensional surface density evolution of a disc is governed by \citep{Cordwell2024}
\begin{align}
    \label{eq:sigma_time}
    \frac{\mathrm{d} \Sigma}{\mathrm{d} t}  = \frac{1}{R} \frac{\mathrm{d} }{\mathrm{d} R} \bigg[& \left( \frac{\mathrm{d} (R^2 \Omega)}{\mathrm{d} R}\right)^{-1} 
    \left\{ \frac{1}{2 \pi} \left(\frac{\mathrm{d} G}{\mathrm{d} R} - \frac{\mathrm{d} \Fdep}{\mathrm{d} R} \right) + \Sigma R \frac{\mathrm{d} (R^2 \Omega)}{\mathrm{d} t} \right\} \bigg],
\end{align}
where all definitions are as given in the main text and 
\begin{equation}
    \label{eq:def_g}
    G \equiv - 2 \pi R^3 \nu \Sigma \frac{\mathrm{d} \Omega}{\mathrm{d} R}
\end{equation}
is the viscous angular momentum flux. Setting all time derivatives in equation (\ref{eq:sigma_time}) to zero and assuming that (i) angular velocities are well approximated as Keplerian and (ii) $\mathrm{d} \Fdep/\mathrm{d}R$ is known as an explicit function of $R$ only (i.e. it is not a function of $\Sigma$), this equation can be solved for the profile of  $\Sigma$:
\begin{align}
    \Sigma(R) &= \frac{1}{3 \pi \nu} \left(C_1 + \frac{\Fdep(R) + C_2}{R^2 \Omega_{\mathrm{K}}} \right),
\end{align}
where $C_1$ and $C_2$ are constants of integration and
\begin{equation}
    \Fdep(R) \equiv \int_{R_i}^{R} \frac{\mathrm{d} \Fdep}{\mathrm{d} R'} d R',
\end{equation}
with $R_i$ being the inner edge of the disc. The constant $C_1$ can be associated with the background viscous mass accretion rate,
\begin{equation}
    C_1 \equiv \dot{M} = \frac{\Sigma_0}{3 \pi \nu}\,,
\end{equation}
whereas $C_2$ represents the angular momentum flux at the inner disc boundary, typically due to angular momentum injected into the disc by the star. For simplicity we will assume that this flux is negligible and set $C_2 = 0$, as in \cite{lynden-bell_evolution_1974, dempsey_pileups_2020}.

The 1D profile surface of the surface density in viscous steady state is then
\begin{equation}
    \label{eq:global_vss}
    \Sigma_{\textrm{Global VSS}}(R) = \frac{1}{3 \pi \nu} \left(\dot{M} + \frac{\Fdep(R)}{R^2 \OmK} \right),
\end{equation}
which is the globally valid form of the solution given in eqn. 19 of \cite{dempsey_pileups_2020}. This form of the 1D profile allows us to directly connect the behavior of $\mathrm{d} \Fdep/\mathrm{d}R$ to the shape of a steady state gap structure.

%%%%%%%%%%%%%%%%%%%%%%%%%%%%%%%%%%%%%%%%%%%%%%%%%%
%%%%%%%%%%%%%%%%%%%%%%%%%%%%%%%%%%%%%%%%%%%%%%%%%%

\section{Additional Plots}
\label{sec:additional_plots}

%%%%%%%%%%%%%%%%%%%%%%%%%%%%%%%%%%%%%%%%%%%%%%%%%%

\begin{figure*}
    \centering
    \includegraphics[width=\linewidth]{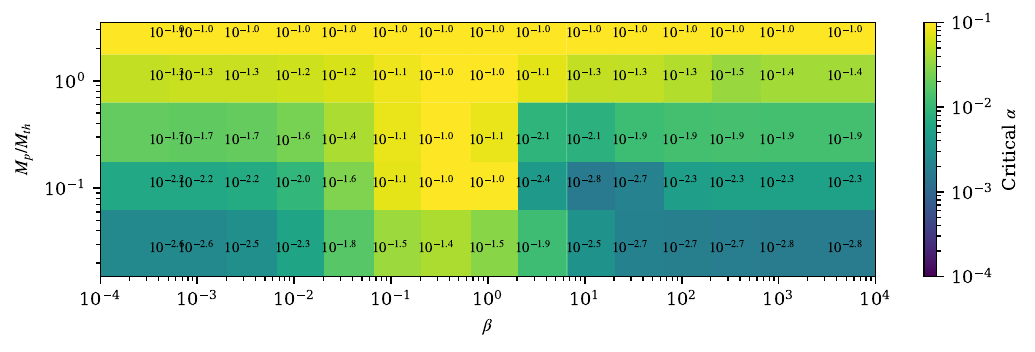}
    \caption{Map of the raw (unsmoothed) critical value of $\alpha$ used in making Figure \ref{fig:critical_alpha}.}
    \label{fig:crit_alpha_exact}
\end{figure*}

\begin{figure*}
    \centering
    \includegraphics[width=\linewidth]{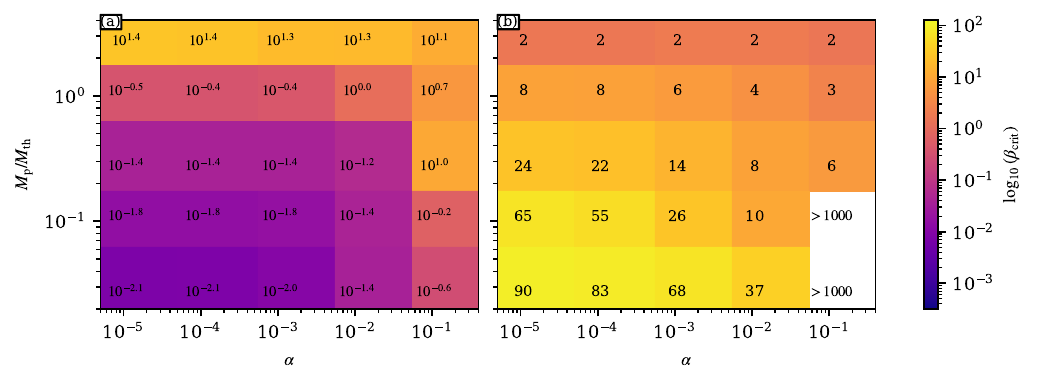}
    \caption{Map of the  raw (unsmoothed) critical values of $\beta$ used in making Figure \ref{fig:critical_beta}.}
    \label{fig:crit_beta_exact}
\end{figure*}

In Figures \ref{fig:crit_alpha_exact} and \ref{fig:crit_beta_exact}, we show the raw data used to produce the contour plots in Figures \ref{fig:critical_alpha} and \ref{fig:critical_beta}. We emphasise however that researchers wishing to use our data to evaluate where viscosity and cooling should be included in their work may be able to get more specific insights using the full $\mathrm{d} \Fdep/\mathrm{d}R$ and surface density evolution database available on zenodo.

%%%%%%%%%%%%%%%%%%%%%%%%%%%%%%%%%%%%%%%%%%%%%%%%%%
%%%%%%%%%%%%%%%%%%%%%%%%%%%%%%%%%%%%%%%%%%%%%%%%%%

\section{Results using the second order potential}
\label{sec:sec_order}

%%%%%%%%%%%%%%%%%%%%%%%%%%%%%%%%%%%%%%%%%%%%%%%%%%

\begin{figure*}
    \centering
    \includegraphics[width=\linewidth]{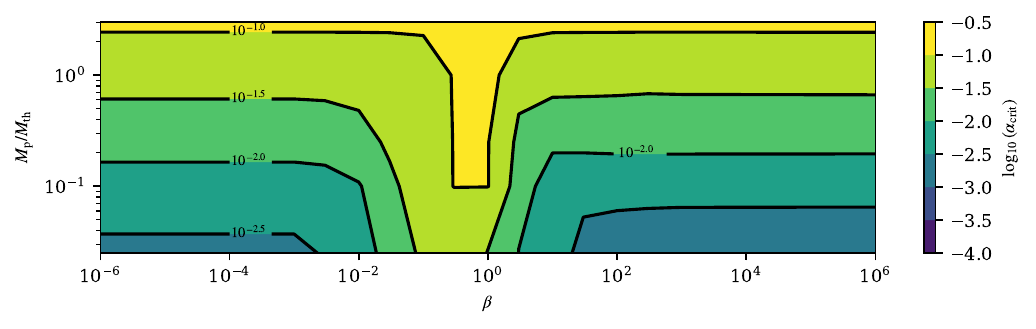}
    \caption{
    Same as Figure \ref{fig:critical_alpha} but made for the second order (Plummer) potential, see Appendix \ref{sec:sec_order}. }
    \label{fig:critical_alpha_2nd}
\end{figure*}
\begin{figure*}
    \centering
    \includegraphics[width=\linewidth]{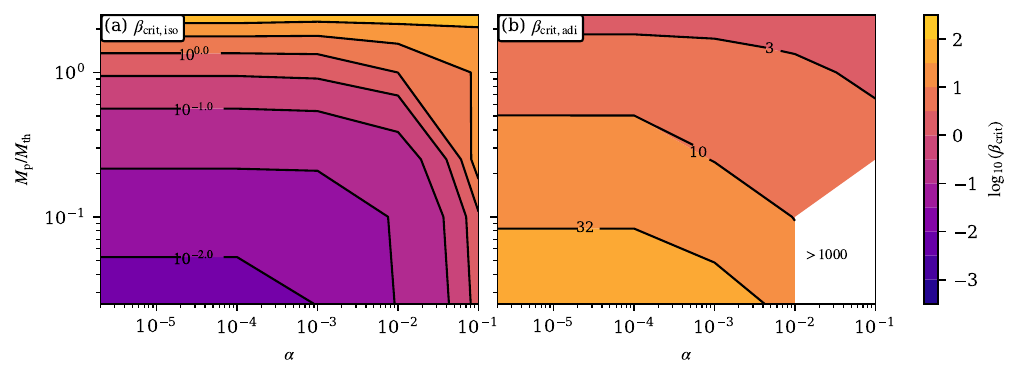}
    \caption{
    Same as Figure \ref{fig:critical_beta} but made for the second order (Plummer) potential, see Appendix \ref{sec:sec_order}.  }
    \label{fig:critical_beta_2nd}
\end{figure*}

As the Bessel-type potential used in the main text (Equation \ref{eq:bessel_const_h}) has not yet been applied widely to studies of planet-disc interactions, we also simulated a version of our grid where we represented the planetary potential using the second-order (Plummer) potential (Equation \ref{eq:plummer}) with a smoothing length of $b = 0.65$. We show results for critical values of $\alpha$ and $\beta$ in Figures \ref{fig:critical_alpha_2nd} and \ref{fig:critical_beta_2nd}. We find that these results are entirely consistent with those using the Bessel-type potential.

%%%%%%%%%%%%%%%%%%%%%%%%%%%%%%%%%%%%%%%%%%%%%%%%%%

% Don't change these lines
\bsp	% typesetting comment
\label{lastpage}
\end{document}